\documentclass[12pt,onecolumn,journal]{IEEEtran}
\usepackage{latexsym}
\usepackage[margin=2.54cm, lmargin=2.54cm]{geometry}
\usepackage{amsfonts}
\usepackage{amsbsy}
\usepackage{amsmath,amssymb}
\usepackage{times}
\usepackage{graphicx}
\usepackage{enumerate}
\usepackage{color}
\usepackage[dvips]{pstcol}
\usepackage{epstopdf}
\usepackage{multicol}
\usepackage{multirow}
\usepackage{cite}
\usepackage{amsmath}
\usepackage{amssymb}
\usepackage{amsfonts}
\usepackage{graphicx}
\usepackage{epsfig}
\usepackage{psfrag}
\usepackage{amsfonts, bm}
\usepackage{epstopdf}
\usepackage{cite}
\usepackage{subfig}
\usepackage{amsthm}
\usepackage{algorithm}
\usepackage{algorithmic}
\usepackage[table,xcdraw]{xcolor}
\usepackage{diagbox}

\usepackage{textcomp,booktabs}
\usepackage{colortbl}
\definecolor{mygray}{gray}{.9}
\definecolor{mypink}{rgb}{.99,.91,.95}
\definecolor{mycyan}{cmyk}{.3,0,0,0}

\usepackage{array}
\makeatletter
\newcommand{\thickhline}{%
    \noalign {\ifnum 0=`}\fi \hrule height 0.8pt
    \futurelet \reserved@a \@xhline
}

\newcolumntype{"}{@{\hskip\tabcolsep\vrule width 0.8pt\hskip\tabcolsep}}

\newtheorem{theorem}{Theorem}
\newtheorem{lemma}{Lemma}

\newtheorem{corollary}{Corollary}
\newtheorem{remark}{Remark}

\newtheorem{Proof}{Proof}

\input epsf
\begin{document}
\title{ \LARGE Iterative Algorithm Induced Deep-Unfolding Neural Networks: Precoding Design for Multiuser MIMO Systems }


\author{ Qiyu Hu, Yunlong Cai, Qingjiang Shi, Kaidi Xu, \\ Guanding Yu, and Zhi Ding
\thanks{
Q. Hu, Y. Cai, K. Xu, and G. Yu are with the College of Information Science and Electronic Engineering, Zhejiang University, Hangzhou 310027, China (e-mail: qiyhu@zju.edu.cn; ylcai@zju.edu.cn; xukaidi13@126.com; yuguanding@zju.edu.cn).
Q. Shi is with the School of Software Engineering, Tongji University, Shanghai, China (e-mail: qing.j.shi@gmail.com).
Z. Ding is with the Department of Electrical and Computer Engineering, University of California,
Davis, CA 95616, USA (e-mail: zding@ucdavis.edu).

}
}

\maketitle
\vspace{-3.3em}
\begin{abstract}
Optimization theory assisted algorithms have received great attention for precoding design in multiuser multiple-input multiple-output (MU-MIMO) systems. Although the resultant optimization algorithms are able to provide excellent performance, they generally require considerable computational complexity, which gets in the way of their practical application in real-time systems. In this work, in order to address this issue, we first propose a framework for deep-unfolding, where a general form of iterative algorithm induced deep-unfolding neural network (IAIDNN) is developed in matrix form to better solve the problems in communication systems. Then, we implement the proposed deep-unfolding framework to solve the sum-rate maximization problem for precoding design in MU-MIMO systems. An efficient IAIDNN based on the structure of the classic weighted minimum mean-square error (WMMSE) iterative algorithm is developed. Specifically, the iterative WMMSE algorithm is unfolded into a layer-wise structure, where a number of trainable parameters are introduced to replace the high-complexity operations in the forward propagation.
To train the network, a generalized chain rule of the IAIDNN is proposed to depict the recurrence relation of gradients between two adjacent layers in the back propagation. 
Moreover, we discuss the computational complexity and generalization ability of the proposed scheme. Simulation results show that the proposed IAIDNN efficiently achieves the performance of the iterative WMMSE algorithm with reduced computational complexity.
\end{abstract}

\IEEEpeerreviewmaketitle

\section{Introduction}
\label{Introduction}
Multiuser multiple-input multiple-output (MU-MIMO) systems have received great attention in wireless communications, since they can dramatically increase the spectrum efficiency \cite{MIMO01,MIMO02,MIMO1,MIMO2,MIMO3}. In order to maximize the spectrum efficiency, a number of efficient iterative precoding design algorithms which are relied on the optimization theory have been proposed for the downlink of MU-MIMO systems \cite{IWFA,SDR,WMMSE,Spectral}.
An iterative water-filling algorithm (IWFA) has been developed for MIMO interference systems in \cite{IWFA}. The authors of \cite{SDR} applied semidefinite relaxation (SDR) to design the transmit precoding for MIMO multicasting systems.
In \cite{WMMSE}, a weighted minimum mean-square error (WMMSE) iterative algorithm has been proposed for precoding design in MU-MIMO systems, where the sum-rate maximization problem is first equivalently transformed into an MMSE problem and then a block coordinate descent (BCD) method is proposed to solve the resultant MMSE problem. 
The authors of \cite{Spectral} proposed an iterative hybrid precoding algorithm based on a novel penalty dual decomposition (PDD) optimization framework. 
Although these iterative precoding algorithms provide approaching theoretical bound performance, they require very high computational complexity due to the large-dimensional matrix inversion and the large number of iterations, especially for the massive MU-MIMO systems in the upcoming 5G communication systems, which hinders their applications in real-time systems. 

Recently, many studies have developed machine learning based algorithms to solve the computationally intensive and time sensitive signal processing tasks for communication. The main idea of this method is to treat the iterative algorithm as a black-box, and learn the mapping between the input and the output by employing the deep neural network (DNN) and the convolutional neural network (CNN) \cite{DL}.
Some representative studies can be found in \cite{LearnOptimize,PowerControl,Towards,Spatial,Graph,Channelesti,CSIfeedback} for different applications, such as resource allocation and channel estimation. 
The first try came from \cite{LearnOptimize} and \cite{PowerControl}, where the authors applied the multi-layer perceptron (MLP) and CNN to approximate the iterative WMMSE algorithm in a multiuser single-input single-output system. The authors of \cite{Towards} proposed an efficient power allocation algorithm by employing unsupervised learning to achieve better performance.
With the aid of underlying topology of wireless networks, several resource allocation schemes based on the spatial convolution and the graph neural network (GNN) have been proposed in \cite{Spatial} and \cite{Graph}, respectively. 
Furthermore, the authors of \cite{Channelesti} and \cite{CSIfeedback} have applied the DNN and CNN in channel estimation and channel state information (CSI) feedback.

However, the black-box based neural network (NNs) suffer from poor interpretability and generalization ability, and have no performance guarantee. The data-driven black-box based NN requires a lot of training samples, which incurs a long training time. To overcome such drawbacks, a number of studies \cite{DeepUnfold,TopicModel,DeepProximal,Interpret,RealTime} have been proposed to unfold the iterations into a layer-wise structure analogous to a NN based on the existing iterative algorithms.
This method is referred to as deep unfolding \cite{UnfoldSurvey} and has a wide range of applications in communications, such as detection and coding \cite{Detect1,Detect2,ISTA,Coding}, resource allocation and channel estimation \cite{AMP,LearnResource,Appli}. 
For MIMO detection, the authors of \cite{Detect1} designed the deep-unfolding NN based on the projected gradient algorithm and a model-driven deep learning NN is developed in \cite{Detect2}. The authors of \cite{ISTA} applied a multi-layer network to approximate the iterative soft-threshold algorithm (ISTA) for sparse coding. In \cite{Coding}, a deep-unfolding based hybrid decoder design for polar code has been proposed.
In addition, an approximate message passing (AMP) inspired NN has been developed in \cite{AMP} for massive MIMO channel estimation. In \cite{LearnResource}, a primal-dual method that learns the parameters of DNN and the primal and dual variables has been proposed to solve the constrained resource allocation problem, and the authors of \cite{Appli} extended it to the scenario of distributed optimization. 

To the best of our knowledge, the deep-unfolding based NNs have not been well investigated for precoding design in MU-MIMO systems.
Moreover, the design of existing deep-unfolding NNs mainly focuses on the optimization of scalar variables. 
In this work, we first propose a general framework for deep-unfolding, where a general form of iterative algorithm induced deep-unfolding neural network (IAIDNN) is developed in matrix form to better solve the problems in communication systems. 
Based on a general iterative algorithm, the structure of IAIDNN is designed in the forward propagation (FP), where a number of trainable parameters are introduced. In the back propagation (BP), the generalized chain rule (GCR) of the IAIDNN in matrix form is proposed, which depicts the recurrence relation of gradients between two adjacent layers. 
The gradients of the trainable parameters in different layers are calculated based on the GCR. It extends the chain rule in DNN, which is the basis of the famous platform ``tensorflow", and we show that the existing chain rule is a special case of our proposed GCR. 

We implement the proposed deep-unfolding framework to solve the sum-rate maximization problem for precoding design in MU-MIMO systems, where an efficient IAIDNN based on the structure of the classic iterative WMMSE algorithm \cite{WMMSE} is developed. Specifically, by integrating the power constraint into the objective function, we obtain an equivalent unconstrained sum-rate maximization problem, the objective function of which is regarded as the loss function in the unsupervised training stage.
To design the IAIDNN, the iterative WMMSE algorithm is unfolded into a layer-wise structure
with a series of matrix multiplication and non-linear operations.
On the one hand, we use much smaller number of iterations, i.e., layers in the IAIDNN, to approximate the iterative WMMSE algorithm, and avoid the matrix inversion to reduce computational complexity. On the other hand, we aim at improving the performance by introducing trainable parameters.
In the FP, we apply the element-wise non-linear function and the first-order Taylor expansion structure of the inverse matrix to approximate the matrix inversion operation. In the BP, we employ the proposed GCR to calculate the gradients of the trainable parameters and update them based on the stochastic gradient descent (SGD) method.
Moreover, we develop a black-box based CNN as a benchmark, and discuss the computational complexity and generalization ability of the proposed schemes. Simulation results show that the proposed IAIDNN significantly outperforms the conventional precoding algorithms and the black-box based CNN, and efficiently achieves the performance of the iterative WMMSE algorithm with reduced computational complexity.
The contributions of this work are summarized as follows.

\begin{itemize}
\item We propose a framework for deep-unfolding, where the general form of IAIDNN is developed in matrix form to better solve the problems in communication systems. To train the IAIDNN, the GCR is proposed to calculate the gradients of the trainable parameters.  

\item We implement the proposed deep-unfolding framework to solve the sum-rate maximization problem for precoding design in MU-MIMO systems. Based on the structure of the iterative WMMSE algorithm, an efficient IAIDNN is developed, where the iterative WMMSE algorithm is unfolded into a layer-wise structure.

\item We analyze the computational complexity and generalization ability of the proposed schemes. Simulation results show that the proposed IAIDNN efficiently achieves the performance of the iterative WMMSE algorithm with reduced computational complexity. The contribution becomes more significant in a massive MU-MIMO system.
\end{itemize}

The paper is structured as follows. Section \ref{Unfolding} proposes a general form of deep-unfolding based framework in matrix form, where the GCR is developed. Section \ref{WMMSEAppro} presents the problem formulation and briefly introduces the  classic WMMSE iterative precoding design algorithm for spectrum efficiency maximization.
Section \ref{AlgoInduce} develops an IAIDNN based on the WMMSE iterative algorithm.
Section \ref{Complexity} presents a black-box based CNN as a benchmark and analyzes the computational complexity and generalization ability of the proposed IAIDNN.
The simulation results are presented in Section \ref{Simulation}
and the conclusion is drawn in Section \ref{Conclusion}.

\emph{Notations:} Scalars, vectors and matrices are respectively denoted by lower case, boldface lower case and boldface upper case letters.
$\mathbf{I}$ represents an identity matrix and $\mathbf{0}$
denotes an all-zero  matrix.
For a matrix $\mathbf{A}$, ${{\bf{A}}^T}$, $\mathbf{A}^*$, ${{\bf{A}}^H}$ and $\|\mathbf{A}\|$ denote its transpose, conjugate, conjugate transpose and Frobenius norm, respectively.
Moreover, ${{\bf{A}}^{-1}}$ denotes the inversion of matrix $\mathbf{A}$, while ${{\bf{A}}^{+}}$ represents the operation that takes the reciprocal of each element in matrix $\mathbf{A}$.
For a vector $\mathbf{a}$, $\|\mathbf{a}\|$ represents its Euclidean norm.
$\mathbb{E}\{ \cdot \}$ denotes the statistical expectation. 
$\textrm{Tr}\{ \cdot \}$ denotes the trace operation.
$|  \cdot  |$ denotes the absolute value of a complex scalar and $\circ$ denotes the element-wise multiplication of two matrices, i.e., Hadmard product.
${\mathbb{C}^{m \times n}}\;({\mathbb{R}^{m \times n}})$ denotes the space of ${m \times n}$ complex (real) matrices.

\section{Proposed Deep-Unfolding Based Framework}
\label{Unfolding}

In this section, we propose a framework for deep-unfolding, where a general form of IAIDNN is developed in matrix form. The GCR is developed to calculate the gradients of the trainable parameters. 

\subsection{ Problem Setup }
The general form of the optimization problem can be formulated as
\begin{equation}
\min\limits_{\mathbf{X}} \quad f(\mathbf{X};\mathbf{Z}) \quad \text{s.t.} \quad \mathbf{X}\in \mathcal{X}, \label{generalform}
\end{equation}
where $f: \mathbb{C}^{m\times n}\mapsto \mathbb{R}$ is a continuous objective function, $\mathbf{X}\in \mathbb{C}^{m\times n}$ is the variable, $\mathcal{X}$ is the feasible region, and $\mathbf{Z}\in \mathbb{C}^{p\times q}$ denotes the random parameter of the problem.

In order to solve Problem \eqref{generalform}, an iterative algorithm can be developed with the following general iteration expression
\begin{equation}
\mathbf{X}^{t}=F_{t}(\mathbf{X}^{t-1};\mathbf{Z}),  \label{generalsolu}
\end{equation}
where $t\in \mathcal{T}\triangleq \{1, 2, \ldots , T\}$ denotes the index of the iteration and $T$ denotes the total number of iterations, and function $F_{t}$ maps the variable $\mathbf{X}^{t-1}$ to the variable $\mathbf{X}^{t}$ at the $t$-th iteration based on the parameter $\mathbf{Z}$.

\subsection{ Forward Propagation }
Based on the structure of the general iteration expression in \eqref{generalsolu}, we introduce the trainable parameter $\bm{\theta}\in \mathbb{C}^{a\times b}$ to reduce the complexity of the iterative algorithm and improve its performance.
Since $\mathbf{Z}$ always turns out to be a random variable, by taking the expectation of $\mathbf{Z}$, Problem \eqref{generalform} can be rewritten as
\begin{equation}
\min\limits_{\mathbf{X}} \quad \mathbb{E}_{\mathbf{Z}}\big\{ f(\mathbf{X};\bm{\theta},\mathbf{Z}) \big\} \quad \text{s.t.} \quad \mathbf{X}\in \mathcal{X}. \label{generalform2}
\end{equation}
Then, the iteration expression shown in \eqref{generalsolu} can be transformed into the following NN,
\begin{equation}
\mathbf{X}^{l}=\mathcal{F}_{l}(\mathbf{X}^{l-1};\bm{\theta}^{l},\mathbf{Z}),  \label{generalnetwork}
\end{equation}
where $l\in \mathcal{L}\triangleq \{1, 2, \ldots , L\}$ is the index of the layer in NN and $L$ denotes the total number of layers, $\mathcal{F}_{l}$ represents the structure of the network in the $l$-th layer, $\mathbf{X}^{l-1}$ and $\mathbf{X}^{l}$ denote the input and output of the $l$-th layer, respectively, $\mathbf{Z}$ is the given parameter or input of the network, and $\bm{\theta}^{l}$ represents the trainable parameter in the $l$-th layer.
Moreover, the objective function $f(\mathbf{X};\mathbf{Z})$ in \eqref{generalform} could serve as the loss function of the NN in \eqref{generalnetwork}. 
Finally, we substitute the output of the network $\mathbf{X}^{L}$  into the objective function $f(\mathbf{X};\mathbf{Z})$ to obtain the final results.

\subsection{Back Propagation}
In order to train parameter $\bm{\theta}^{l}$, we need to calculate the gradient of the objective function $f(\mathbf{X};\bm{\theta},\mathbf{Z})$ with respect to $\bm{\theta}^{l}$, and perform the BP to update $\bm{\theta}^{l}$. To depict the recurrence relation of gradients between two adjacent layers, we propose the GCR in matrix form in Theorem \ref{theorem Chain rule}. In Remark \ref{remarkBP}, we compare the proposed GCR with the existing chain rule of the DNN, which is a special case of the proposed GCR. The proof of Theorem \ref{theorem Chain rule} is presented in Appendix \ref{Appendix 0}.

\begin{theorem} [\textbf{GCR in matrix form}]   \label{theorem Chain rule} 
	Recall the general structure of NN presented in \eqref{generalnetwork}: $\mathbf{X}^{l}=\mathcal{F}_{l}(\mathbf{X}^{l-1};\bm{\theta}^{l},\mathbf{Z})$.
	The recurrence relation of the gradients from $\mathbf{X}^{l}$ to $\mathbf{X}^{l-1}$ in adjacent layers can be written as
    \begin{equation} \label{Chain rule}
	\begin{aligned}
	\textrm{Tr} ( \mathbf{G}^{l}d\mathbf{X}^{l} )
	=\textrm{Tr} \bigg( \mathbf{G}^{l}\mathbf{A}^{l} \big( \mathbf{B}^{l}\circ ( \mathbf{C}^{l}d\mathbf{X}^{l-1}\mathbf{E}^{l}) \big)\mathbf{F}^{l} \bigg)
	&\overset{\eqref{tracepro}}{=} \textrm{Tr} \bigg( \mathbf{E}^{l}\big( (\mathbf{F}^{l}\mathbf{G}^{l}\mathbf{A}^{l})\circ (\mathbf{B}^{l})^{T} \big) \mathbf{C}^{l}d\mathbf{X}^{l-1} \bigg),
	\end{aligned}
	\end{equation}
	where $\mathbf{G}^{l}$ and $\mathbf{G}^{l-1}$ are the gradients of $\mathbf{X}^{l}$ and $\mathbf{X}^{l-1}$, respectively, and other matrices, e.g., $\mathbf{A}^{l}$, are related to the structure of the NN, i.e., $\bm{\theta}^{l}$ and $\mathbf{Z}$. Note that the NN always applies the element-wise non-linear function, then we introduce $\circ$ to denote the element-wise multiplication, i.e., Hadmard product. Thus, we obtain the GCR in matrix form as
	\begin{equation} \label{matrix chain rule}
	\mathbf{G}^{l-1}=\mathbf{E}^{l}\big( ( \mathbf{F}^{l}\mathbf{G}^{l}\mathbf{A}^{l})\circ (\mathbf{B}^{l})^{T} \big) \mathbf{C}^{l}.
	\end{equation}
\end{theorem}

\begin{remark} \label{remarkBP}
	Recall the structure in existing DNNs: $\mathbf{x}^{l}=\varphi(\mathbf{z}^{l})$, where $\mathbf{z}^{l}=\mathbf{W}^{l}\mathbf{x}^{l-1}+\mathbf{b}^{l}$. Note that $\mathbf{W}^{l}$ and $\mathbf{b}^{l}$ represent the trainable weight and offset matrix in the $l$-th layer, respectively, $\mathbf{z}^{l}$ and $\mathbf{x}^{l}$ denote the input and output of $\varphi$ in the $l$-th layer, respectively, and $\varphi$ denotes the element-wise non-linear  function.
	Then, its chain rule in scalar form is:
	\begin{equation}   \label{scalar chain rule}
	\mathbf{g}^{l-1}=( (\mathbf{W}^{l})^{T}\mathbf{g}^{l}  )\circ \varphi'(\mathbf{z}^{l}),
	\end{equation}
	where $\mathbf{g}^{l}$ denotes the gradient of $\mathbf{x}^{l}$ in the $l$-th layer. By comparing \eqref{matrix chain rule} with \eqref{scalar chain rule},
	if we let $\mathbf{F}=\mathbf{W}^{T}$, $\mathbf{B}=\varphi'(\mathbf{z})^{T}$, and $\mathbf{A}=\mathbf{E}=\mathbf{C}=\mathbf{I}$, where $\mathbf{I}$ represents the identity matrix, then \eqref{matrix chain rule} equals to \eqref{scalar chain rule}. Thus, \eqref{scalar chain rule} is a special case of \eqref{matrix chain rule}.
\end{remark}

Then, to illustrate how to use Theorem \ref{theorem Chain rule}, we provide a specific NN with quadratic structure and element-wise non-linear function. Based on \eqref{matrix chain rule} in Theorem \ref{theorem Chain rule}, we derive its recurrence relation of gradients in adjacent layers.

\begin{corollary} \label{Case}
    Specifically, if the network has the following  quadratic structure:
    \begin{equation}
    \mathbf{X}^{l} = \mathbf{\bar{A}}\mathbf{X}^{l-1}\mathbf{\bar{B}}\mathbf{X}^{l-1}\mathbf{\bar{C}} + \varphi(\mathbf{\bar{A}}\mathbf{X}^{l-1}\mathbf{\bar{B}}\mathbf{X}^{l-1}\mathbf{\bar{C}})\mathbf{\bar{D}},
    \end{equation}
    where $\varphi:\mathbb{C}^{m\times n}\mapsto \mathbb{C}^{m\times n}$ is an element-wise non-linear function, $\mathbf{X}^{l}$ is the output matrix in the $l$-th layer, while the others, e.g., $\mathbf{\bar{A}}$, are constant matrices or trainable parameters, and we omit $l$ for these matrices for clarity. The loss function is $f(\mathbf{X}), f:\mathbb{C}^{m\times n}\mapsto \mathbb{R}$.  Then the differential with respect to $\mathbf{X}^{l-1}$ is given by
    \begin{equation}
    \begin{aligned}
    \textrm{Tr}\bigg\{ \mathbf{G}^{l}d\mathbf{X}^{l} \bigg \}
    &\overset{\eqref{Chain rule}}{=} \textrm{Tr}\bigg\{ \bigg( \mathbf{\bar{B}}\mathbf{X}^{l-1}\mathbf{\bar{C}}\mathbf{G}^{l} \big( \mathbf{\bar{D}}\circ \varphi'(\mathbf{\bar{A}}\mathbf{X}^{l-1}\mathbf{\bar{B}}\mathbf{X}^{l-1}\mathbf{\bar{C}})^{T}+\mathbf{I}  \big) \mathbf{\bar{A}}  \\
    &\quad \quad \quad \quad + \mathbf{\bar{C}}\mathbf{G}^{l} \big( \mathbf{\bar{D}}\circ \varphi'(\mathbf{\bar{A}}\mathbf{X}^{l-1}\mathbf{\bar{B}}\mathbf{X}^{l-1}\mathbf{\bar{C}})^{T}+\mathbf{I}  \big) \mathbf{\bar{A}}\mathbf{X}^{l-1}\mathbf{\bar{B}} \bigg) d\mathbf{X}^{l-1}  \bigg\},
    \end{aligned}
    \end{equation}
    where $\varphi'$ denotes the element-wise derivative corresponding to $\varphi$, 
    $\mathbf{G}^{l}$ is the gradient of $\mathbf{X}^{l}$ in the $l$-th layer.
\end{corollary}

The quadratic structure is commonly used in the NN design and we will apply the results in Corollary \ref{Case} to design the IAIDNN in the following section. 
Based on Theorem \ref{theorem Chain rule} and Corollary \ref{Case}, the BP process can be summarized as below. Firstly, the gradient of $\mathbf{X}^{L}$ in the last layer, i.e., $\mathbf{G}^{L}$, is derived by differentiating $f(\mathbf{X})$ with respect to $\mathbf{X}^{L}$, which is the output of the NN. Then, based on the GCR presented in Theorem  \ref{theorem Chain rule}, the gradient in each layer, i.e., $\{ \mathbf{G}^{l}, l\in \mathcal{L} \}$ is obtained. Finally, the gradient of trainable parameter $\bm{\theta}^{l}$ is calculated based on $\mathbf{G}^{l}$. 

\section{ Iterative WMMSE Precoding Design Algorithm }
\label{WMMSEAppro}


In this section, we briefly introduce the classic WMMSE iterative precoding design algorithm. 

\subsection{Problem Formulation}
The iterative WMMSE algorithm is one of the most representative precoding design algorithms for maximizing the system sum-rate \cite{WMMSE}.
We consider a downlink MU-MIMO system consisting a BS which is equipped with $N_{t}$ transmit antennas and $K$ users, each of which is equipped with $N_{r,k}$ receive antennas, where $k \in \mathcal{K}\triangleq \{1, 2, \ldots , K\}$.
Let $\mathbf{V}_{k}\in \mathbb{C}^{N_{t}\times d}$ represent the precoding matrix that the BS applies to process the transmit signal vector $\mathbf{s}_{k}\in \mathbb{C}^{d\times 1}$ for user $k\in \mathcal{K}$, then we have the precoded data vector as 
\begin{equation}
\mathbf{x}=\sum\limits_{k=1}^{K} \mathbf{V}_{k}\mathbf{s}_{k}.   \notag
\end{equation}
Here, we assume that $\mathbf{s}_k$ is with zero mean and $ \mathbb{E}[\mathbf{s}_{k}\mathbf{s}_{k}^{H}]=\mathbf{I} $, and the symbols sent by different users are independent from each other. Then, the received signal vector $\mathbf{y}_k \in \mathbb{C}^{N_{r,k}\times 1}$ at user $k$ is given by
\begin{equation}
\mathbf{y}_{k}=\mathbf{H}_{k}\mathbf{V}_{k}\mathbf{s}_{k} + \sum\limits_{m=1, m\neq k}^{K}\mathbf{H}_{k} \mathbf{V}_{m}\mathbf{s}_{m} + \mathbf{n}_{k}, \quad \forall k\in \mathcal{K},  \notag
\end{equation}
where $\mathbf{H}_{k}\in \mathbb{C}^{N_{r,k}\times N_{t}}$ denotes the MIMO channel matrix from the BS to user $k$ and $\mathbf{n}_{k}\in \mathbb{C}^{N_{r,k}\times 1}$ represents the additive noise, which is modeled as a circularly symmetric complex Gaussian random vector with zero-mean and correlation matrix
$\mathcal{CN}(0,\sigma_{k}^{2}\mathbf{I})$ and $\sigma^2_k$ denotes the average noise power at user $k$. 

We aim at optimizing the precoding matrices to maximize the system sum-rate subject to the transmit power constraint. Thus, the problem is formulated as
\begin{subequations} \label{sum-rate max}
\begin{eqnarray}
& \max\limits_{ \{\mathbf{V}_{k}\} } &\sum\limits_{k=1}^{K}\omega_k \log \det\bigg( \mathbf{I} + \mathbf{H}_{k}\mathbf{V}_{k}\mathbf{V}_{k}^{H}\mathbf{H}_{k}^{H}
\big(  \sum\limits_{m\neq k} \mathbf{H}_{k}\mathbf{V}_{m}\mathbf{V}_{m}^{H}\mathbf{H}_{k}^{H} + \sigma_{k}^{2} \mathbf{I}    \big)^{-1}   \bigg) \label{originobj} \\
&\text{s.t.}  & \sum\limits_{k=1}^{K} \textrm{Tr}(\mathbf{V}_{k}\mathbf{V}^{H}_{k})\leq P_{T},   \label{power constraint}
\end{eqnarray}
\end{subequations}
where the weight $\omega_k$ represents the priority of user $k$ in the system, $P_{T}$ denotes the total transmit power budget at the BS, and constraint \eqref{power constraint} reflects the power constraint.

\subsection{Iterative WMMSE Precoding Design}
It has been proved in \cite{WMMSE} that the MMSE problem in \eqref{sum-MSE min} shown below is equivalent to the sum-rate maximization problem in \eqref{sum-rate max}, in the sense that the optimal solution $\{\mathbf{V}_k\}$ for the two problems are identical
\begin{subequations} \label{sum-MSE min}
\begin{eqnarray}
& \min\limits_{ \{ \mathbf{W}_{k}, \mathbf{U}_{k}, \mathbf{V}_{k} \} } &\sum\limits_{k=1}^{K} \omega_k \big( \textrm{Tr}( \mathbf{W}_{k}\mathbf{E}_{1,k} )-\log \det (\mathbf{W}_{k}) \big) \label{transferobj} \\
&\text{s.t.}  & \sum\limits_{k=1}^{K} \textrm{Tr}(\mathbf{V}_{k}\mathbf{V}^{H}_{k})\leq P_{T} ,
\end{eqnarray}
\end{subequations}
where $\mathbf{U}_{k}$ and $\mathbf{W}_{k}$ are introduced auxiliary variables, and 
\begin{equation}
\mathbf{E}_{1,k}\triangleq ( \mathbf{I}-\mathbf{U}_{k}^{H}\mathbf{H}_{k}\mathbf{V}_{k} )( \mathbf{I}-\mathbf{U}_{k}^{H}\mathbf{H}_{k}\mathbf{V}_{k} )^{H}
+\sum\limits_{m\neq k} \mathbf{U}_{k}^{H}\mathbf{H}_{k}\mathbf{V}_{m}\mathbf{V}_{m}^{H}\mathbf{H}_{k}^{H}\mathbf{U}_{k} + \sigma_{k}^{2} \mathbf{U}_{k}^{H} \mathbf{U}_{k}. \notag
\end{equation}

For the convenience of the IAIDNN design, we integrate the term $\frac{1}{P_T}\!\! \sum\limits_{k=1}^K \!\! \textrm{Tr}(\! \mathbf{V}_k \!\! \mathbf{V}^H_k )$ into the objective function \eqref{originobj} and consider the following unconstrained sum-rate maximization problem
\begin{equation}
\max\limits_{ \{\mathbf{V}_{k}\} } \sum\limits_{k=1}^{K} \omega_k \log \det\bigg( \mathbf{I} + \mathbf{H}_{k}\mathbf{V}_{k}\mathbf{V}_{k}^{H}\mathbf{H}_{k}^{H}
\big(  \sum\limits_{m\neq k} \mathbf{H}_{k}\mathbf{V}_{m}\mathbf{V}_{m}^{H}\mathbf{H}_{k}^{H} + \frac{\sigma_{k}^{2}}{P_{T}} \sum\limits_{n=1}^K \textrm{Tr}(\mathbf{V}_{n}\mathbf{V}^{H}_{n})\mathbf{I}    \big)^{-1}    \bigg).     \label{objlossfunc0}
\end{equation}
Then, we have the following lemma,
\begin{lemma}  \label{twosolution}
The optimal solution $\mathbf{V}_{k}^{\star}$ of the original problem in \eqref{sum-rate max} and the optimal solution $\mathbf{V}_{k}^{\star\star}$ of the transformed unconstrained problem in \eqref{objlossfunc0} satisfy the following relation: $\mathbf{V}_{k}^{\star}=\alpha\mathbf{V}_{k}^{\star\star}$, where $\alpha= \frac{ \sqrt{P_{T}} }{ \big(\sum\limits_{k=1}^K \textrm{Tr}(\mathbf{V}_{k}^{\star\star}(\mathbf{V}_{k}^{\star\star})^{H}) \big)^{\frac{1}{2} } } $ is a scaling factor. 
\end{lemma}

\begin{Proof} [Proof of Lemma \ref{twosolution}]
It is readily seen that $\mathbf{V}^{\star}_k$ always makes constraint \eqref{power constraint} meet equality. By substituting $\alpha \mathbf{V}^{\star\star}_k$ into \eqref{objlossfunc0} and power constraint \eqref{power constraint}, we generate the maximum value of the objective function and meet the equality of the constraint, respectively. Therefore, $\alpha \mathbf{V}^{\star\star}_k$ is the optimal solution of problem \eqref{sum-rate max}.
\end{Proof}

Therefore, we can solve the unconstrained optimization problem \eqref{objlossfunc0} instead of problem \eqref{sum-rate max}, followed by the scaling operation. Furthermore, based on the relation between problem \eqref{sum-rate max} and \eqref{sum-MSE min}, the resultant more tractable problem \eqref{objlossfunc0} can be equivalently transformed into the following unconstrained MMSE problem
\begin{equation}
\min\limits_{ \{ \mathbf{W}_{k}, \mathbf{U}_{k}, \mathbf{V}_{k} \} } \sum\limits_{k=1}^{K} \omega_k \big( \textrm{Tr}( \mathbf{W}_{k}\mathbf{E}_{2,k} )-\log \det (\mathbf{W}_{k}) \big), \label{transobj}
\end{equation}
where $\mathbf{E}_{2,k}$ is given by
\begin{equation}
\mathbf{E}_{2,k}\triangleq ( \mathbf{I}-\mathbf{U}_{k}^{H}\mathbf{H}_{k}\mathbf{V}_{k} )( \mathbf{I}-\mathbf{U}_{k}^{H}\mathbf{H}_{k}\mathbf{V}_{k} )^{H}
+ \sum\limits_{m\neq k} \mathbf{U}_{k}^{H}\mathbf{H}_{k}\mathbf{V}_{m}\mathbf{V}_{m}^{H}\mathbf{H}_{k}^{H}\mathbf{U}_{k} + \frac{ \sum\limits_{n} \textrm{Tr}(\mathbf{V}_{n}\mathbf{V}^{H}_{n}) }{P_{T}} \sigma_{k}^{2}\mathbf{U}_{k}^{H} \mathbf{U}_{k}.  \notag
\end{equation}

\begin{figure}[t]
\begin{centering}
\includegraphics[width=0.99\textwidth]{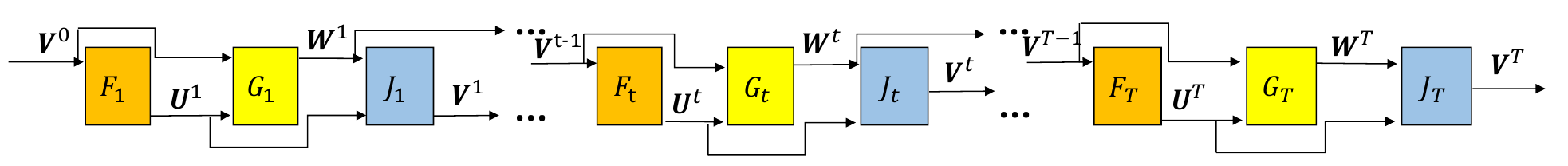}
\par\end{centering}
\caption{The diagram of the iterative WMMSE precoding design algorithm.}
\label{iterative}
\end{figure}

Based on \cite{WMMSE}, a BCD type iterative algorithm can be developed to solve Problem \eqref{transobj}. It converges to a stationary point of the original Problem \eqref{sum-rate max}. The details of this algorithm are shown in \textbf{Algorithm \ref{WMMSE algorithm}}, where we omit the iteration index $t$ for clarity.

\begin{algorithm*}[t]\caption{ Iterative WMMSE precoding design algorithm  } \label{WMMSE algorithm}
\begin{algorithmic}
\begin{small}
\STATE Initialize $\{ \mathbf{V}_{k} \}$ to satisfy $ \sum\limits_{k=1}^K  \textrm{Tr}(\mathbf{V}_{k}\mathbf{V}^{H}_{k})\leq P_{T} $. Set the tolerance of accuracy $\epsilon$, the maximum iteration number $I_{\max}$, and the current iteration index $t = 0$. 
\REPEAT 
\STATE 1. Update $\mathbf{U}_{k}$: $ \mathbf{U}_{k} = \mathbf{A}_{k}^{-1} \mathbf{H}_{k}\mathbf{V}_{k} $, where $ \mathbf{A}_{k} = \frac{\sigma_{k}^{2}}{P_{T}} \sum\limits_{k=1}^K \textrm{Tr}(\mathbf{V}_{k}\mathbf{V}^{H}_{k})\mathbf{I} + \sum\limits_{m=1}^K \mathbf{H}_{k}\mathbf{V}_{m}\mathbf{V}_{m}^{H}\mathbf{H}_{k}^{H}, \quad \forall k, $   \\
\STATE 2. Update $\mathbf{W}_{k}$: $ \mathbf{W}_{k} = \mathbf{E}_{k}^{-1} $, where $ \mathbf{E}_{k} = \mathbf{I}-\mathbf{U}_{k}^{H}\mathbf{H}_{k}\mathbf{V}_{k},  \quad  \forall k $,     \\
\STATE 3. Update $\mathbf{V}_{k}$: \!\! $ \mathbf{V}_{k} \!\!=\!\! \omega_k \mathbf{B}^{-1}\mathbf{H}_{k}^{H}\mathbf{U}_{k}\mathbf{W}_{k} $, where $ \mathbf{B} \!\!=\!\! \sum\limits_{k=1}^K \!\! \frac{\sigma_{k}^{2}}{P_{T}} \textrm{Tr}(\omega_k \mathbf{U}_{k}\mathbf{W}_{k}\mathbf{U}_{k}^{H})\mathbf{I} +\! \sum\limits_{m=1}^K \!\!\! \omega_m \mathbf{H}_{m}^{H}\mathbf{U}_{m}\mathbf{W}_{m}\mathbf{U}_{m}^{H}\mathbf{H}_{m},   \forall k  $,  \\
4. $t=t+1$. 
\UNTIL
The objective function converges or $t \geq I_{\max}$. Scale $\{ \mathbf{V}_{k} \}$ to meet the transmit power constraint. 
\end{small}
\end{algorithmic}
\end{algorithm*}
 
Comparing Problem \eqref{transobj} with Problem \eqref{generalform}, we have the following identification
\begin{equation}
\mathbf{X}\equiv \{ \mathbf{W}_{k}, \mathbf{U}_{k},\mathbf{V}_{k}, \forall k \in \mathcal{K} \}, \quad \mathbf{Z}\equiv  \{ \mathbf{H}_{k}, \omega_k, \sigma_{k}, P_{T}, \forall k \in \mathcal{K} \} .  \notag
\end{equation}
Corresponding to line 1-3 in \textbf{Algorithm \ref{WMMSE algorithm}}, the iterative WMMSE algorithm can be rewritten in the general form of iterative algorithm presented in Section \ref{Unfolding}, which is given by
\begin{subequations} \label{sum-MSE iter}
\begin{eqnarray}
& & \mathbf{U}^{t}=F_{t}(\mathbf{V}^{t-1}),  \\
& & \mathbf{W}^{t}=G_{t}(\mathbf{U}^{t}, \mathbf{V}^{t-1} ),  \\
& & \mathbf{V}^{t}=J_{t}(\mathbf{U}^{t}, \mathbf{W}^{t} ),
\end{eqnarray}
\end{subequations}
where $F_{t}$, $G_{t}$, and $J_{t}$ are iterative mapping functions at the $t$-th iteration. The flowchart of the iterative process for the WMMSE precoding design algorithm is presented in Fig. \ref{iterative}.

\section{ Proposed IAIDNN for Precoding Design }
\label{AlgoInduce}

In this section, we introduce the proposed IAIDNN based on the classic iterative WMMSE precoding design algorithm.

\subsection{Architecture of the IAIDNN and Its Forward Propagation}
\begin{figure}[t]
\begin{centering}
\includegraphics[width=0.99\textwidth]{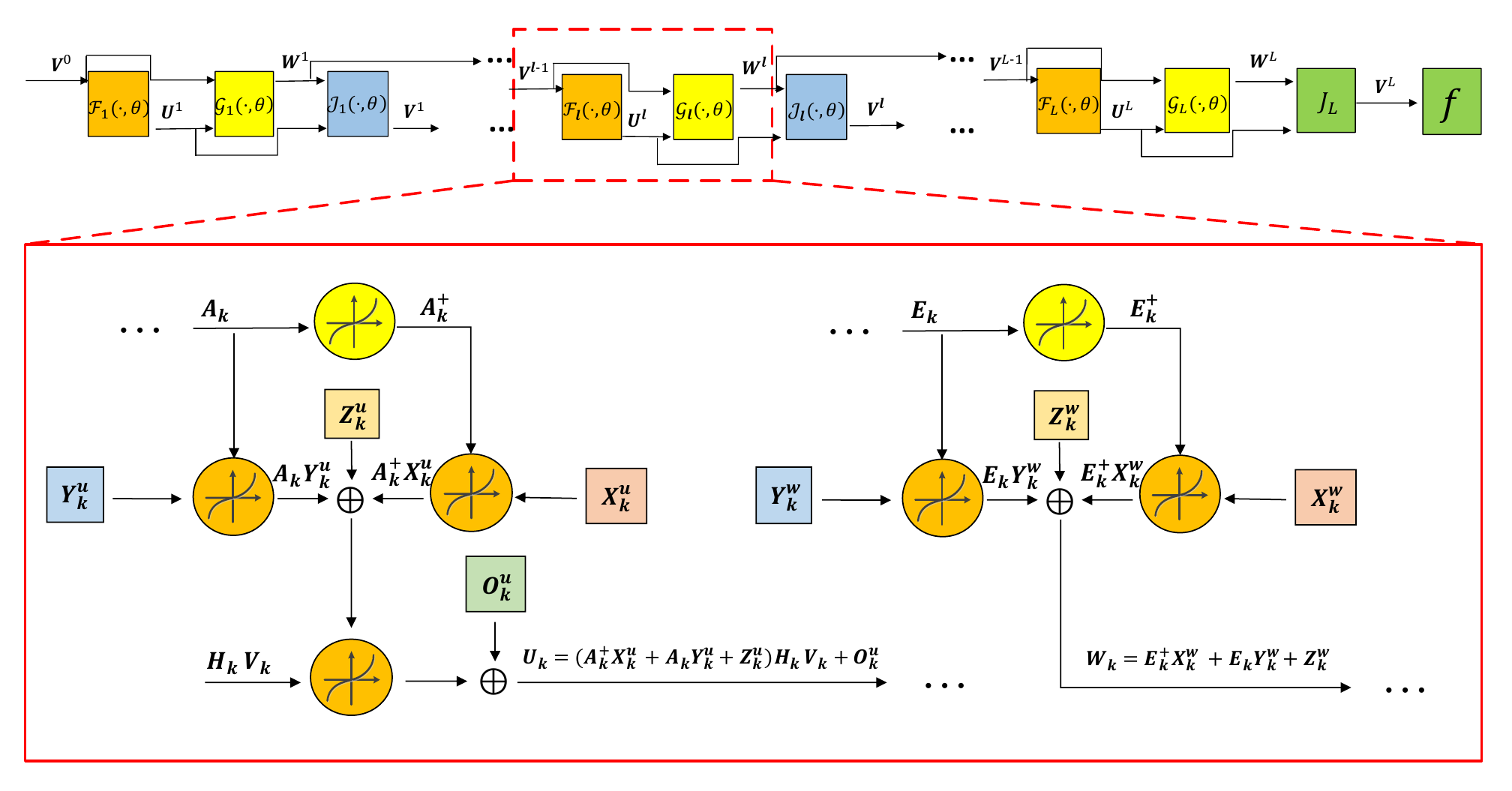}
\par\end{centering}
\caption{The architecture of the IAIDNN for precoding design. The red frame shows the detailed structure in each layer, where the circles represent the non-linear operation and the squares denote the trainable parameters.}
\label{UWframe}
\end{figure}

We define two kinds of non-linear operations: (i) Multiplication of matrix variables; (ii) The element-wise non-linear operation that takes the reciprocal of each element in the diagonal of matrix $\mathbf{A}$ while sets the non-diagonal elements to be $0$, i.e., denoted as $\mathbf{A}^{+}$. 
Since the matrix inversion $\mathbf{A}^{-1}$ has high computational complexity, we approximate it by employing the combination of the following two structures with lower complexity. 

\begin{itemize}
\item Firstly, we apply the structure $\mathbf{A}^{+}\mathbf{X}$ with the element-wise non-linear operation $\mathbf{A}^{+}$ and trainable matrix parameter $\mathbf{X}$, where $\mathbf{X}$ is introduced to improve the performance.
Note that when $\mathbf{A}$ is a diagonal matrix, we have $\mathbf{A}^{-1}=\mathbf{A}^{+}$. Moreover, we observe that the diagonal elements of the matrices tend to be much larger than the non-diagonal elements in the iterative WMMSE algorithm. Thus, $\mathbf{A}^{+}$ might be a good approximation of $\mathbf{A}^{-1}$ here with lower complexity.
\item Secondly, by recalling the first-order Taylor expansion structure of the inverse matrix $\mathbf{A}^{-1}$ at $\mathbf{A}_0$: $\mathbf{A}^{-1}=2\mathbf{A}_{0}^{-1}-\mathbf{A}_{0}^{-1}\mathbf{A}\mathbf{A}_{0}^{-1}$, we use the structure $\mathbf{A}\mathbf{Y}
+ \mathbf{Z}$ with trainable matrix parameters $\mathbf{Y}$ and $\mathbf{Z}$ to approximate $\mathbf{A}^{-1}$. 
\end{itemize}
	
Thus, we apply  $\mathbf{A}^{+}\mathbf{X}+\mathbf{A}\mathbf{Y}+\mathbf{Z}$ to approximate the matrix inversion $\mathbf{A}^{-1}$.
Note that $\{ \mathbf{X}_{k}^{u,l+1}, \\ \mathbf{Y}_{k}^{u,l+1},  \mathbf{Z}_{k}^{u,l+1} \}$, $\{ \mathbf{X}_{k}^{w,l+1}, \mathbf{Y}_{k}^{w,l+1}, \mathbf{Z}_{k}^{w,l+1} \}$, and $\{ \mathbf{X}_{k}^{v,l+1},  \mathbf{Y}_{k}^{v,l+1},  \mathbf{Z}_{k}^{v,l+1} \}$ are introduced trainable parameter sets to approximate the inversion of matrix variables $\mathbf{U}_{k}^{l+1}$, $\mathbf{W}_{k}^{l+1}$, and $\mathbf{V}_{k}^{l+1}$ in the $(l+1)$-th layer, respectively, and $\{ \mathbf{O}_{k}^{u,l+1}, \mathbf{O}_{k}^{v,l+1} \}$ denote the trainable offsets.
The structure of the network can be designed as
\begin{subequations}  \label{networkUWV}
\begin{eqnarray}
& & \mathbf{U}_{k}^{l+1} = \bigg( (\mathbf{A}_{k}^{l})^{+}\mathbf{X}_{k}^{u,l+1} + \mathbf{A}_{k}^{l}\mathbf{Y}_{k}^{u,l+1}
+ \mathbf{Z}_{k}^{u,l+1}  \bigg) \mathbf{H}_{k}\mathbf{V}_{k}^{l} + \mathbf{O}_{k}^{u,l+1} ,  \\
& & \mathbf{W}_{k}^{l+1} = (\mathbf{E}_{k}^{l+1})^{+}\mathbf{X}_{k}^{w,l+1} + \mathbf{E}_{k}^{l+1}\mathbf{Y}_{k}^{w,l+1} + \mathbf{Z}_{k}^{w,l+1}  ,   \\
& & \mathbf{V}_{k}^{l+1} = \bigg( (\mathbf{B}^{l+1})^{+}\mathbf{X}_{k}^{v,l+1} + \mathbf{B}^{l+1}\mathbf{Y}_{k}^{v,l+1} + \mathbf{Z}_{k}^{v,l+1}  \bigg) \omega_k \mathbf{H}_{k}^{H}\mathbf{U}_{k}^{l+1}\mathbf{W}_{k}^{l+1} + \mathbf{O}_{k}^{v,l+1},
\end{eqnarray}
\end{subequations}
where
\begin{subequations} \label{networkABE}
\begin{eqnarray}
& & \mathbf{A}_{k}^{l} \triangleq   \frac{\sigma^{2}_{k}}{P_{T}} \sum\limits_{k=1}^K \textrm{Tr}(\mathbf{V}_{k}^{l}(\mathbf{V}^{l}_{k})^{H})\mathbf{I} +
\sum\limits_{m=1}^K \mathbf{H}_{k}\mathbf{V}^{l}_{m}(\mathbf{V}^{l}_{m})^{H}\mathbf{H}_{k}^{H} ,   \\
& & \mathbf{B}^{l+1} \triangleq  \sum\limits_{k=1}^K \frac{\sigma^{2}_{k}}{P_{T}} \textrm{Tr}(\omega_k \mathbf{U}^{l+1}_{k}\mathbf{W}^{l+1}_{k}(\mathbf{U}^{l+1}_{k})^{H})\mathbf{I} + \sum\limits_{m=1}^K \omega_m \mathbf{H}_{m}^{H}\mathbf{U}^{l+1}_{m}\mathbf{W}^{l+1}_{m}(\mathbf{U}^{l+1}_{m})^{H}\mathbf{H}_{m} ,  \\
& & \mathbf{E}_{k}^{l+1} \triangleq \mathbf{I}-(\mathbf{U}^{l+1}_{k})^{H}\mathbf{H}_{k}\mathbf{V}^{l}_{k}  .
\end{eqnarray}
\end{subequations}

The architecture of the proposed IAIDNN is shown in Fig. \ref{UWframe}.
For the simplicity of notation, we drop the index $l$ for the matrix variables.
Since the dimension of $\mathbf{U}_k$ and $\mathbf{W}_k$ is much smaller than that of $\mathbf{V}_k$, it is better to treat $\mathbf{U}_k$ and $\mathbf{W}_k$ as the output of the IAIDNN. Thus, we apply the iterative expression of $\mathbf{V}_k$ in the WMMSE algorithm in the last layer, i.e., the module $J_{L}$ in Fig. \ref{UWframe}, which is given by
\begin{equation} \label{iterativeV}
\mathbf{V}_{k} = \bigg(  \sum\limits_{k=1}^K \frac{\sigma_{k}^{2}}{P_{T}} \textrm{Tr}(\omega_k
\mathbf{U}_{k}\mathbf{W}_{k}\mathbf{U}_{k}^{H})\mathbf{I} + \sum\limits_{m=1}^K \omega_m \mathbf{H}_{m}^{H}\mathbf{U}_{m}\mathbf{W}_{m}\mathbf{U}_{m}^{H}\mathbf{H}_{m} \bigg)^{-1} \omega_k \mathbf{H}_{k}^{H}\mathbf{U}_{k}\mathbf{W}_{k} ,   \forall k.
\end{equation}
Since the channel matrices $ \mathbf{H}_{k} $ are random variables, we take the expectation of $\mathbf{H}_{k}$ and modify the objective function in \eqref{objlossfunc0} into
\begin{equation}
\max\limits_{ \{\mathbf{V}_{k}\} } \sum\limits_{k=1}^{K} \mathbb{E}_{\mathbf{H}_{k}} \bigg\{ \! \omega_k \log \det\bigg( \! \mathbf{I} \!+ \mathbf{H}_{k}\mathbf{V}_{k}\mathbf{V}_{k}^{H}\mathbf{H}_{k}^{H}
\big( \!\! \sum\limits_{m\neq k}\! \mathbf{H}_{k}\mathbf{V}_{m}\mathbf{V}_{m}^{H}\mathbf{H}_{k}^{H} \!+\! \frac{\sigma_{k}^{2}}{P_{T}} \sum\limits_{k} \textrm{Tr}(\mathbf{V}_{k}\mathbf{V}^{H}_{k})\mathbf{I}    \big)^{-1}  \!  \bigg) \! \bigg\}.    \label{objlossfunc}
\end{equation}
Then, \eqref{objlossfunc} could be regarded as the loss function of the NN, i.e., the module $f$ in Fig. \ref{UWframe}.
Moreover, to avoid gradient explosion, we normalize each $\mathbf{V}_{k}$ by $P_{T}$ at the end of each layer, i.e., $\frac{ \sqrt{P_{T}} }{ \big( \sum\limits_{k} \textrm{Tr}(\mathbf{V}_{k}^{l+1}(\mathbf{V}_{k}^{l+1})^{H}) \big)^{\frac{1}{2}} }\mathbf{V}_{k}^{l+1}$, to satisfy the power constraint in \eqref{power constraint}.

\subsection{Generalized Chain Rule and Back Propagation}
Firstly, by substituting \eqref{iterativeV} into the objective function \eqref{objlossfunc0}, we can calculate the gradient with respect to $\mathbf{U}_k^{L}$ and $\mathbf{W}_k^{L}$ for each sample in the last layer, i.e., $\{ \mathbf{G}_{k}^{u,L}, \mathbf{G}_{k}^{w,L} \}$. The detailed gradients are presented in Appendix \ref{Appendix A}.

Secondly, based on the GCR in matrix form shown in Theorem \ref{theorem Chain rule}, we obtain the recurrence relation from the gradients of $\{ \mathbf{U}_{n}, \mathbf{W}_{n}, \mathbf{V}_{n}, n\in \mathcal{K} \}$ in the $(l+1)$-th layer, i.e., $\{\! \mathbf{G}_{n}^{u,l+1},\! \mathbf{G}_{n}^{w,l+1},\!  \mathbf{G}_{n}^{v,l+1} \! \}$ to those in the $l$-th layer, i.e., $\{ \mathbf{G}_{n}^{u,l}, \mathbf{G}_{n}^{w,l}, \mathbf{G}_{n}^{v,l} \}$. 
The details of calculating the gradients $\{ \mathbf{G}_{n}^{u,l}, \\ \mathbf{G}_{n}^{w,l}, \mathbf{G}_{n}^{v,l}, \forall l\in \mathcal{L}, n\in \mathcal{K} \}$ in each layer are presented in Appendix \ref{Appendix B}.

Then, based on the structure of the IAIDNN in \eqref{networkUWV} and $\{ \mathbf{G}_{n}^{u,l},   \mathbf{G}_{n}^{w,l}, \mathbf{G}_{n}^{v,l}, \forall l\in \mathcal{L}, n\in \mathcal{K} \}$ in \eqref{gradientGW}-\eqref{gradientGV}, the gradients of trainable parameters are calculated as follows
\begin{equation} \label{gradientXYZO}
\begin{aligned}
& \nabla_{\mathbf{X}_{k}^{u,l+1}}f = \mathbf{H}_{k}\mathbf{V}_{k}^{l}\mathbf{G}_{k}^{u,l+1}(\mathbf{A}_{k}^{l})^{+},  \quad
\nabla_{\mathbf{X}_{k}^{w,l+1}}f = \mathbf{G}_{k}^{w,l+1}(\mathbf{E}_{k}^{l+1})^{+}, \quad
\nabla_{\mathbf{Z}_{k}^{w,l+1}}f = \mathbf{G}_{k}^{w,l+1},      \\
& \nabla_{\mathbf{Y}_{k}^{u,l+1}}f = \mathbf{H}_{k}\mathbf{V}_{k}^{l}\mathbf{G}_{k}^{u,l+1}\mathbf{A}_{k}^{l},  \quad
\nabla_{\mathbf{Y}_{k}^{w,l+1}}f = \mathbf{G}_{k}^{w,l+1}\mathbf{E}_{k}^{l+1},      \quad
\nabla_{\mathbf{Z}_{k}^{u,l+1}}f = \mathbf{H}_{k}\mathbf{V}_{k}^{l}\mathbf{G}_{k}^{u,l+1},   \\
& \nabla_{\mathbf{X}_{k}^{v,l+1}}f = (\mathbf{H}_{k})^{H}\mathbf{U}_{k}^{l+1}\mathbf{W}_{k}^{l+1}\mathbf{G}_{k}^{v,l+1}(\mathbf{B}^{l+1})^{+},     \quad
\!\! \nabla_{\mathbf{O}_{k}^{v,l+1}}f = \mathbf{G}_{k}^{v,l+1},  \quad
\nabla_{\mathbf{O}_{k}^{u,l+1}}f = \mathbf{G}_{k}^{u,l+1},    \\
& \nabla_{\mathbf{Y}_{k}^{v,l+1}}f = (\mathbf{H}_{k})^{H}\mathbf{U}_{k}^{l+1}\mathbf{W}_{k}^{l+1}\mathbf{G}_{k}^{v,l+1}\mathbf{B}^{l+1},  \quad \quad
\nabla_{\mathbf{Z}_{k}^{v,l+1}}f = (\mathbf{H}_{k})^{H}\mathbf{U}_{k}^{l+1}\mathbf{W}_{k}^{l+1}\mathbf{G}_{k}^{v,l+1}.
\end{aligned}
\end{equation}

\begin{figure}[t]
\begin{centering}
\includegraphics[width=0.99\textwidth]{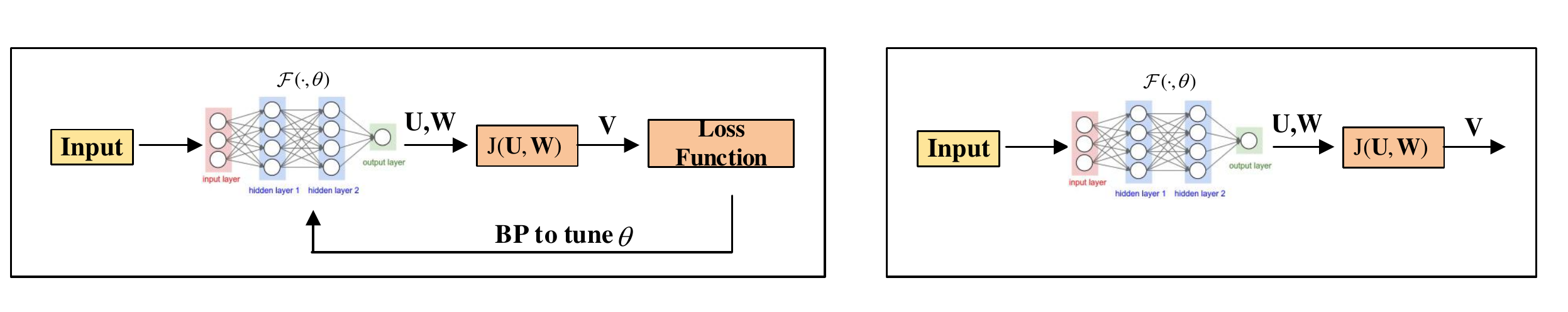}
\par\end{centering}
\caption{Training stage and testing stage.}
\label{trainingtest}
\end{figure}

We apply the gradient descent method to train the IAIDNN, i.e., $(\mathbf{X}_{k}^{u,l})^{m+1} = (\mathbf{X}_{k}^{u,l})^{m} + \sigma_{m} \nabla_{(\mathbf{X}_{k}^{u,l})^{m}}f$, where $\sigma_{m}$ denotes the step size and $\nabla_{(\mathbf{X}_{k}^{u,l})^{m}}f$ denotes the gradient of $\mathbf{X}_{k}^{u,l}$ in layer $l$ at the $m$-th iteration in the training stage.
The gradients of the trainable parameters are presented in \eqref{gradientXYZO}, where $m$ is omitted for clarity.
We choose the step size $\sigma_{m}$ based on \cite{ALiu}, which satisfies the following conditions: $\sigma_{m}$ is decreasing with the iteration number $m$, $\sigma_{m}\in (0,1]$, $\sigma_{m}\rightarrow 0$, $\sum_{m}\sigma_{m}\rightarrow \infty$, and $\sum_{m}(\sigma_{m})^{2}< \infty$, e.g., $\sigma_{m}=m^{-\alpha}, 0<\alpha<1$.
The trainable parameters are initialized randomly, and $\mathbf{V}^{0}_{k}$ is initialized by using the zero-forcing precoder.
The process of training stage and testing stage is presented in Fig. \ref{trainingtest}, and the detailed training procedures of the IAIDNN are presented in \textbf{Algorithm \ref{algorithm-induced}}.

\begin{algorithm*}[t]\caption{Training procedures of the IAIDNN} \label{algorithm-induced}
\begin{algorithmic}
\begin{small}
\STATE Given the training set $\mathcal{H}$. Set the number of layers $L$, the batch size $N$, the tolerance of accuracy $\epsilon$, the maximum iteration number $I_{\max}$, and the current iteration index of the training stage $m = 0$.  
\REPEAT
\STATE 1. \textbf{Forward propagation}: Select a group of samples $\{\mathbf{H}_{k},\forall k\}$ from the training set and initialize $\{\mathbf{V}_{k}^{0},\forall k\}$. Compute $\{ \mathbf{U}_{k}^{l},\mathbf{W}_{k}^{l}, l=1,2,\cdots, L, \forall k\}$ and $\{ \mathbf{V}_{k}^{l}, l=1,2,\cdots, L-1, \forall k \}$ based on  \eqref{networkUWV}-\eqref{networkABE}.   \\
\STATE 2. Compute $\{\mathbf{V}_{k}^{L}, \forall k\}$ based on \eqref{iterativeV} in the last layer. Then plug $\{\mathbf{V}_{k}^{L}, \forall k\}$ into the loss function and obtain its value.  \\
\STATE 3. \textbf{Back propagation}: Firstly, compute the gradients with respect to variables $\mathbf{U}_k^{L}$ and $\mathbf{W}_k^{L}$ in the last layer based on Appendix \ref{Appendix A}. Secondly, compute the gradients of $\{\mathbf{U}_{k}^{l},\mathbf{W}_{k}^{l},\mathbf{V}_{k}^{l}, l=L-1,\cdots, 2,1,  \forall k\}$ according to \eqref{gradientGW}-\eqref{gradientGV} in Appendix \ref{Appendix B}. Finally, compute the gradients of trainable parameters $\{  \mathbf{X}_{k}^{u,l},  \mathbf{Y}_{k}^{u,l},  
\mathbf{Z}_{k}^{u,l},
\mathbf{O}_{k}^{u,l}  \}$, $\{ \mathbf{X}_{k}^{w,l},  \mathbf{Y}_{k}^{w,l},  \mathbf{Z}_{k}^{w,l}  \}$, and $\{ \mathbf{X}_{k}^{v,l},  \mathbf{Y}_{k}^{v,l},  \mathbf{Z}_{k}^{v,l},  \mathbf{O}_{k}^{v,l} \}$ based on \eqref{gradientXYZO}. \\
\STATE 4. \textbf{Update trainable parameters}: Repeat steps 1-3 for $N$ times and compute the average gradients of trainable parameters in a batch. Then, apply mini-batch SGD to update the trainable parameters.      \\
\STATE 5. $m=m+1$. 
\UNTIL
The loss function in the validation set  converges or $m \geq I_{\max}$.
\end{small}
\end{algorithmic}
\end{algorithm*}

\section{Computational Complexity and Generalization Ability}
\label{Complexity}

In this section, we present a black-box based CNN as a benchmark. Furthermore, the parameter dimension, computational complexity, and the generalization ability of the proposed schemes are analyzed.

\subsection{Conventional Black-Box Based CNN}
\begin{figure}[t]
\begin{centering}
\includegraphics[width=0.99\textwidth]{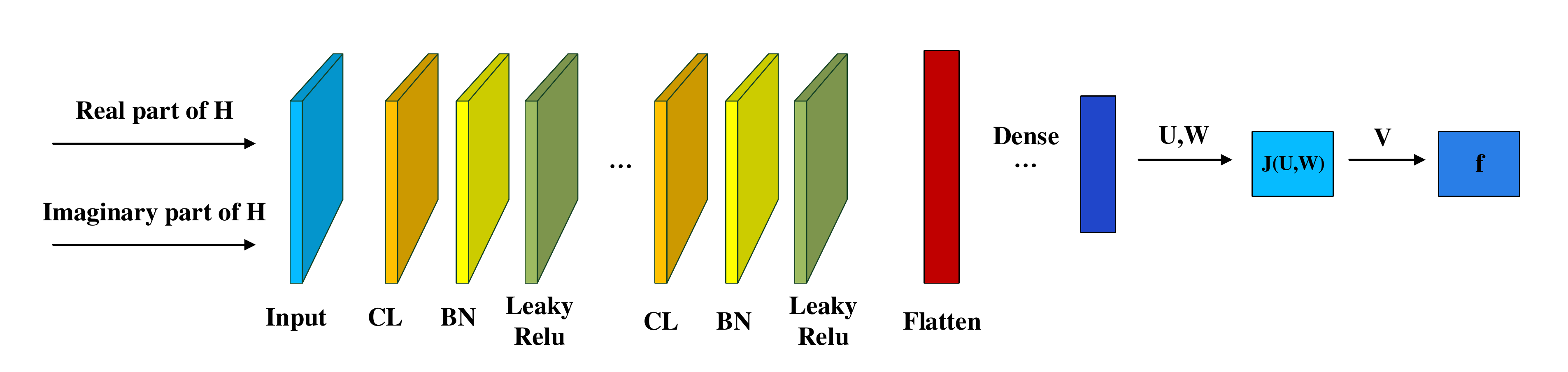}
\par\end{centering}
\caption{The architecture of the black-box based CNN.}
\label{blackbox}
\end{figure}

Based on \cite{LearnOptimize}, we introduce the design of black-box based CNN, which is employed to compare with the proposed IAIDNN as a benchmark.

The architecture of the black-box based CNN is presented in Fig. \ref{blackbox}. Its input is the channel matrix $\mathbf{H}\triangleq [\mathbf{H}_{1}^{T}, \mathbf{H}_{2}^{T}, \cdots ,\mathbf{H}_{k}^{T}]^{T}$, which passes through the convolutional layer (CL), the batch normalization (BN), and the non-linear function in serial. The process repeats for a number of times. Then, it comes through the fully connected (FC) layer, where we apply the flatten and dropout techniques. In particular, we adopt leaky ReLU as the non-linear function, i.e., $y=x$ if $x\geq 0$ and $y=\frac{x}{a}$ if $x<0$, where $a$ is a constant. The outputs of the CNN are auxiliary variables $\mathbf{U}_{k}$ and $\mathbf{W}_{k}$ instead of $\mathbf{V}_{k}$, since the learning effect of the low dimensional variables $\mathbf{U}_{k}$ and $\mathbf{W}_{k}$ is better than that of the $\mathbf{V}_{k}$ with higher dimension. Subsequently, we plug $\mathbf{U}_{k}$ and $\mathbf{W}_{k}$ into the iterative expression in \eqref{iterativeV} to calculate $\mathbf{V}_{k}$. Finally, $\mathbf{V}_{k}$ is substituted into the loss function, which ends the forward propagation. The BP is processed by the platform ``tensorflow" automatically.
We employ the unsupervised learning to improve the performance of the black-box based CNN. Then, the training stage can be divided into the following two parts,
\begin{itemize}
\item Supervised learning stage: Firstly, we apply $ \sum\limits_{k=1}^K \big( \| \mathbf{U}_{k}-\hat{\mathbf{U}}_{k}\|^{2} +\| \mathbf{W}_{k}-\hat{\mathbf{W}}_{k}\|^{2} \big)$ as the loss function, where $\hat{\mathbf{U}}_{k}$ and $\hat{\mathbf{W}}_{k}$ are labels produced by the iterative WMMSE algorithm.
\item Unsupervised learning stage: After applying the supervised learning several times, we use \eqref{objlossfunc} as the loss function.
\end{itemize}
The unsupervised learning is terminated when the loss function converges in the validation set.

\subsection{Parameter Dimension and Computational Complexity}

Then, we discuss the parameter dimension and computational complexity of the proposed IAIDNN, the conventional black-box based CNN, and the iterative WMMSE algorithm. 

\subsubsection{Parameter Dimension}
The parameter dimension of the IAIDNN corresponds to the dimension of $\{ \mathbf{X}_{k}^{u,l}, \mathbf{Y}_{k}^{u,l}, \mathbf{Z}_{k}^{u,l}, \mathbf{O}_{k}^{u,l} \}$, $\{ \mathbf{X}_{k}^{w,l}, \mathbf{Y}_{k}^{w,l}, \mathbf{Z}_{k}^{w,l} \}$, and $\{ \mathbf{X}_{k}^{v,l},  \mathbf{Y}_{k}^{v,l},  \mathbf{Z}_{k}^{v,l},  \mathbf{O}_{k}^{v,l} \}$. 
Then, the parameter dimension in each layer is given by $(3N_{r}^{2}+3d^{2}+3N_{t}^{2}+dN_{r}+dN_{t})K$. Since there is no parameters $\{ \mathbf{X}_{k}^{v,l}, \mathbf{Y}_{k}^{v,l}, \mathbf{Z}_{k}^{v,l}, \mathbf{O}_{k}^{v,l} \}$ in the last layer, the total dimension of parameters is $LK(3N_{r}^{2}+3d^{2}+dN_{r})+(L-1)K(3N_{t}^{2}+dN_{t})$, where $L$ denotes the number of layers.

The parameter dimension in the black-box based CNN is given by $\sum\limits_{l=1}^{L-2}S_{l}^{2}C_{l-1}C_{l} +KN_{r}N_{t} \\ C_{L-2}C_{out}$, where $S_{l}$ and $C_{l}$ represent the size of the convolution kernel and the number of channel at the $l$-th layer in CL, respectively. $C_{out}$ denotes the output size of the FC layer. The first and second terms represent the parameter dimension in the CL and FC layer, respectively. 
We set $S_{l}=5$, $C_{l}=32, \forall l$, and $C_{out}=1024$.

\subsubsection{Computational Complexity}
The computational complexity of the classic iterative WMMSE algorithm is given by $\mathcal{O}\big(L_{w} ( K^{2}N_{t}N_{r}^{2}+K^{2}N_{t}^{2}N_{r}+KN_{t}^{3}+KN_{r}^{3} )\big)$, where $L_{w}$ denotes the number of iterations.

The computational complexity of the proposed IAIDNN in the inference stage is given by $\mathcal{O}\big(L_{a} ( K^{2}N_{t}N_{r}^{2}+K^{2}N_{t}^{2}N_{r}+KN_{t}^{2.37}+KN_{r}^{2.37} )\big)$, where $L_{a} (L_{a}<<L_{w})$ denotes the number of layers. The computational complexity of the IAIDNN is lower than that of the iterative WMMSE algorithm in two aspects:
\begin{itemize}
   \item The number of layers in the IAIDNN is much less, i.e., $L_{a}<<L_{w}$.
   \item The iterative WMMSE algorithm requires the matrix inversion operation, the computational complexity of which is $\mathcal{O}(n^{3})$. In comparison, the proposed IAIDNN only requires matrix multiplication with computational complexity $\mathcal{O}(n^{2.37})$. 
\end{itemize}

Moreover, the computational complexity of the black-box based CNN in the inference stage is $\mathcal{O}\big( \sum\limits_{l=1}^{L-2}M_{l}^{2}S_{l}^{2} C_{l-1}C_{l}+KN_{r}N_{t}C_{L-2}C_{out} + ( K^{2}N_{t}N_{r}^{2}+K^{2}N_{t}^{2}N_{r}+KN_{t}^{3}+KN_{r}^{3} ) \big)$, where $M_{l}=(\frac{P_{l}-S_{l}+2*P_{a}}{S_{t}}+1)$ denotes the output size in the $l$-th layer. Note that $P_{l}$, $P_{a}$, and $S_{t}$ represent the input size, the padding number, and the stride of the $l$-th layer, respectively.

In the training stage, since the proposed IAIDNN applies the closed-form gradients shown in \eqref{gradientXYZO} to update the parameters, it is more efficient with much shorter training time compared to the conventional black-box based CNN, especially in the unsupervised training stage.

\subsection{Analysis of Generalization Ability}

\subsubsection{Generalization Ability}
When the IAIDNN with given $(N_{t_{0}}, N_{r_{0}}, K_{0})$ is trained, it can be  straightforwardly transferred to the scenario with the same parameters $N_{t_{1}}$ and $N_{r_{1}}$ but smaller $K_{1}$, i.e., $( N_{t_{1}}=N_{t_{0}}, N_{r_{1}}=N_{r_{0}}, K_{1}<K_{0} )$, rather than training a new network. In the inference stage, we only need to enter $\{ \mathbf{H}_{k},k\leqslant K_{1} \}$ and $\{ \mathbf{H}_{k}=\mathbf{0}, K_{1}<k\leqslant K_{0} \}$ as the input.
In the case of $N_{t_{1}}<N_{t_{0}}$ and $N_{r_{1}}<N_{r_{0}}$, we set the corresponding column and row vectors in $\mathbf{H}_{k}$ to be $\mathbf{0}$. For example, for the case $N_{t_{0}}=32, N_{t_{1}}=16, N_{r_{0}}=N_{r_{1}}=2, K_{0}=10, K_{1}=5$, we can transfer the trained model from the system $(N_{t_{0}}, N_{r_{0}}, K_{0})$ to the system $(N_{t_{1}}, N_{r_{1}}, K_{1})$. For each sample, we only need to enter $\{ \mathbf{H}_{k},k\leqslant 5 \}$ and $\{ \mathbf{H}_{k}=\mathbf{0}, 5<k\leqslant 10 \}$ as the input, meanwhile add $16$ zero column vectors to $\{ \mathbf{H}_{k},k\leqslant 5 \}$.

\subsubsection{Straightforward Extension}
For clarity, we have assumed that the transmit power $P_{T}$ and noise $\sigma_{k}$ are given in the proposed IAIDNN. It is a straightforward extension to treat them as inputs of the IAIDNN, together with the channel matrices $\{\mathbf{H}_{k}\}$. For example, we can assume $P_{T}\sim \mathcal{N}(a_1,b_1)$, $\sigma_{k}\sim \mathcal{N}(a_2,b_2)$, where $\mathcal{N}(a,b)$ denotes the Gaussian distribution with mean $a$ and variance $b$. 

The proposed IAIDNN is also applicable to the robust precoding algorithm design in the presence of CSI errors and the scenario of multicell systems \cite{WMMSE}. One can easily extend the proposed IAIDNN to these cases by considering the CSI error statistics and slightly adjusting the objective function.

\section{Simulation Results}
\label{Simulation}
In this section, we verify the effectiveness of the proposed IAIDNN by simulation results.

\subsection{Simulation Setup}
In the simulation, we employ the uncorrelated MIMO fading channel model, i.e., the elements in $\mathbf{H}_k$ are generated based on the complex Gaussian distribution $\mathcal{CN}(0,1)$. We set SNR $=20$ dB, and assume that all users are equipped with $N_{r}=2$ receive antennas. We set $N=10$ as the batch size and $L=7$ as the number of layers in the proposed IAIDNN.
For each setup, we run $5,000$ channel matrices in the test data set and take the average of their values of loss function to approximate their expectation in the testing stage. We run the iterative WMMSE  algorithm $30$ times with different initial values and then retain the best result as its performance, which is an approximation of the global optimal solution.
The percentages of the IAIDNN in the table are calculated via dividing the values of sum-rate achieved by the IAIDNN by those of the iterative WMMSE algorithm. The percentages of the black-box based CNN in the table are calculated in the same way.

\subsection{Sum-Rate Performance}

\begin{table*}[t]
	\caption{The sum-rate performance of the analyzed schemes for $N_{t}=8$ and $N_{t}=16$. }
	\begin{center}
		\centering
		\begin{tabular}{c"cccc"cccc}
			\thickhline
			\centering
            \textbf{\# of transmit antennas $(N_{t})$}   & \multicolumn{4}{c"}{8}  &  \multicolumn{4}{c}{16}         \\ \hline
            \rowcolor{mygray}
			\textbf{\# of user $(K)$}   & $1$ & $2$ & $3$ & $4$ & $2$ & $4$ & $6$ & $8$     \\ \hline
			\textbf{WMMSE (bits/s/Hz)} & $13.13$ & $22.12$ & $27.74$ & $31.82$  & $25.93$ & $43.34$ & $53.66$ & $58.83$      \\
            \rowcolor{mygray}
			\textbf{IAIDNN} & $99.34\%$ & $99.15\%$ & $97.36\%$ & $91.35\%$  & $99.59\%$ & $99.11\%$ & $97.67\%$ & $  92.13\%$ \\
            \textbf{Black-box} & $93.76\%$ & $92.09\%$ & $90.27\%$ & $81.85\%$ & $93.68\%$ & $92.39\%$ & $89.32\%$ & $80.56\%$  \\
			\thickhline
		\end{tabular}
	\end{center}
	\label{Nt8}
\end{table*}

\begin{table*}[t]
	\caption{The sum-rate performance of the analyzed schemes for $N_{t}=32$.}
	\begin{center}
		\centering
		\begin{tabular}{cccccccc}
            \rowcolor{mygray}
			\thickhline
			\centering
			\textbf{\# of users $(K)$}   & $4$ & $6$ & $8$ & $10$  & $12$ & $14$ & $16$     \\
			\textbf{WMMSE (bits/s/Hz)} & $51.53$ & $70.38$ & $85.95$ & $97.69$  & $104.78$ & $108.55$ & $113.22$      \\
            \rowcolor{mygray}
			\textbf{IAIDNN} & $99.84\%$ & $99.69\%$ & $99.28\%$ & $99.13\%$ & $98.76\%$ & $97.39\% $ & $92.63\% $  \\
            \textbf{Black-box} & $93.52\%$ & $92.86\%$ & $91.57\%$ & $90.03\%$  & $88.16\% $ & $ 85.24\% $ & $80.39\% $ \\
			\thickhline
		\end{tabular}
	\end{center}
	\label{Nt32}
\end{table*}

\begin{table*}[t]
	\caption{The sum-rate performance of the analyzed schemes for $N_{t}=64$.}
	\begin{center}
		\centering
		\begin{tabular}{ccccccc}
            \rowcolor{mygray}
			\thickhline
			\centering
			\textbf{\# of users $(K)$}   & $5$ & $10$ & $15$ & $20$  & $25$ & $30$      \\
			\textbf{WMMSE (bits/s/Hz)} & $71.04$ & $123.15$ & $164.71$ & $194.72$  & $208.96$ & $216.92$     \\
            \rowcolor{mygray}
			\textbf{IAIDNN} & $99.91\%$ & $99.82\%$ & $99.76\%$ & $99.11\%$ & $98.88\%$ & $97.58\% $   \\
            \textbf{Black-box} & $93.56\%$ & $92.83\%$ & $92.02\%$ & $90.38\%$  & $87.98\%$ & $82.14\%$ \\
			\thickhline
		\end{tabular}
	\end{center}
	\label{Nt64}
\end{table*}

\begin{table*}[t]
	\caption{The sum-rate performance of the analyzed schemes for $N_{t}=128$.}
	\begin{center}
		\centering
		\begin{tabular}{ccccccc}
			\thickhline
			\rowcolor{mygray}
			\textbf{\# of users $(K)$}   & $10$ & $20$ & $30$ & $40$  & $50$ & $60$      \\
			\textbf{WMMSE (bits/s/Hz)} & $139.15$ & $244.03$ & $326.76$ & $386.21$  & $412.82$ & $417.64$     \\
            \rowcolor{mygray}
			\textbf{IAIDNN} & $99.87\%$ & $99.67\%$ & $99.32\%$ & $99.03\%$ & $98.68\%$ & $97.87\% $   \\
            \textbf{Black-box} & $93.58\%$ & $92.76\%$ & $91.58\%$ & $89.93\%$  & $87.31\% $ & $83.89\%$ \\
			\thickhline
		\end{tabular}
	\end{center}
	\label{Nt128}
\end{table*}

\begin{table*}[t]
	\caption{The sum-rate performance of the analyzed schemes for $N_{t}=256$.}
	\begin{center}
		\centering
		\begin{tabular}{cccccccc}
            \rowcolor{mygray}
			\thickhline
			\centering
			\textbf{\# of users $(K)$}    & $20$ & $30$ & $40$  & $50$ & $60$ & $70$ & $80$    \\
			\textbf{WMMSE (bits/s/Hz)}  & $279.56$ & $389.31$ & $487.56$  & $575.32$ & $652.59$ & $718.41$ & $772.07 $  \\
            \rowcolor{mygray}
			\textbf{IAIDNN}  & $99.86\%$ & $99.73\%$ & $99.51\%$ & $99.29\%$ & $99.05\% $ & $98.79\%$ & $98.33\%$ \\
            \textbf{Black-box} & $93.23\%$ & $92.97\%$ & $92.28\%$ & $91.45\%$  & $89.96\% $ & $88.03\% $ & $85.86\% $  \\
			\thickhline
		\end{tabular}
	\end{center}
	\label{Nt256}
\end{table*}

In the following, we evaluate the sum-rate performance of different schemes versus the number of users $K$ and the number of transmit antennas $N_{t}$. 
From the results shown in Table \ref{Nt8}, \ref{Nt32}, \ref{Nt64}, \ref{Nt128} and \ref{Nt256}, we observe that the sum-rate performance achieved by the proposed IAIDNN is close to that of the iterative WMMSE algorithm. The gap between the performance of the proposed IAIDNN and that of the iterative WMMSE algorithm increases with $K$. We can also see that the proposed  IAIDNN outperforms the conventional black-box based CNN, and the gap between the performance of conventional black-box based CNN and that of the IAIDNN also increases with $K$.
Besides, the performance gap between the conventional black-box based CNN and the proposed IAIDNN becomes much larger when $K\times N_{r}$ approaches $N_t$, and the performance of the conventional black-box based CNN deteriorates severely.
It is mainly because the difference of the elements in $\mathbf{V}_{k}$ tends to increase with $K$, i.e., become close to either $0$ or $1$, which makes the NNs more difficult to learn satisfactory results.

\begin{figure}[!t]
\centering
	\subfloat[Different numbers of batch size.]{\centering \scalebox{0.50}{\includegraphics{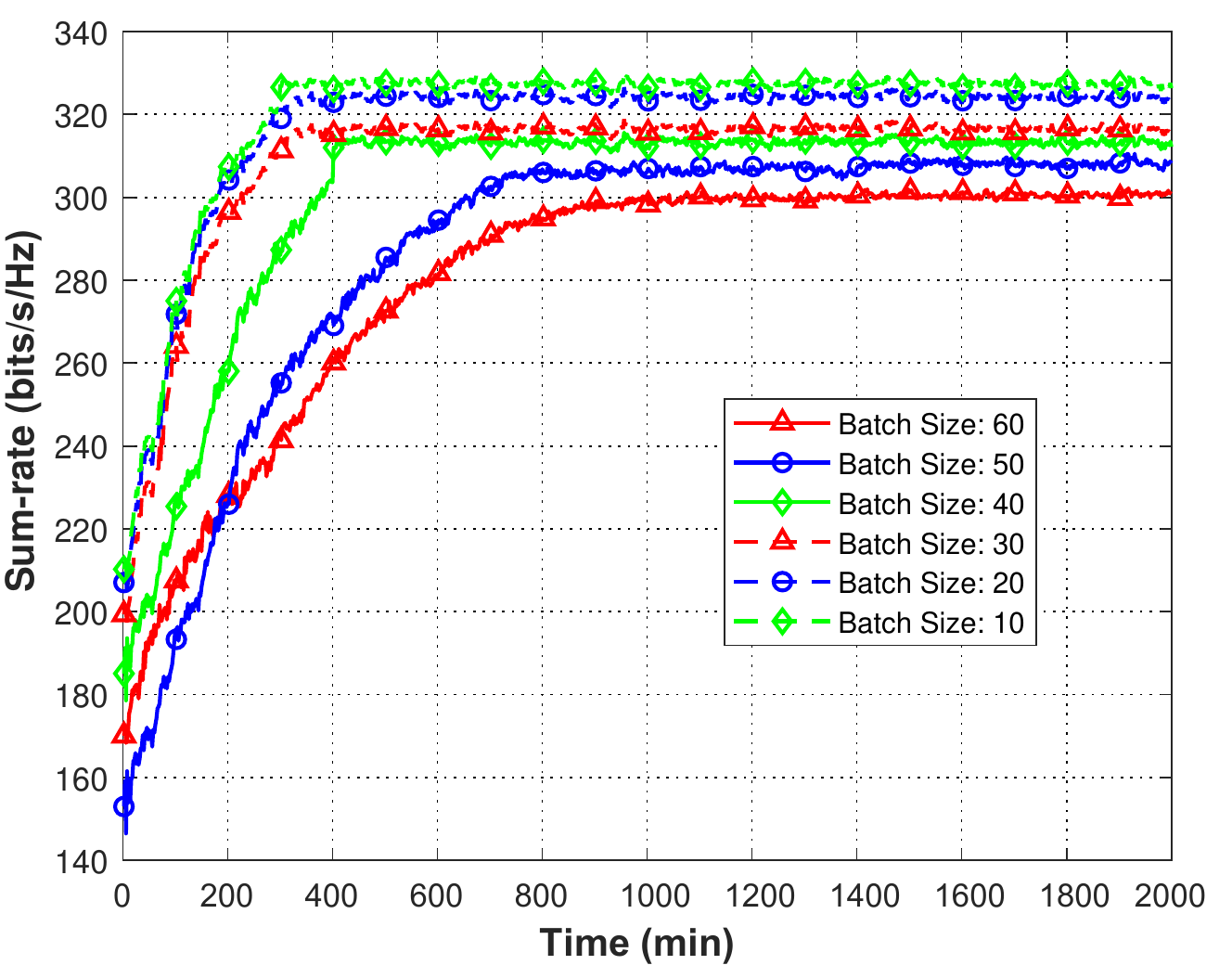}} }
	\subfloat[Different choices of learning rate.]{\centering \scalebox{0.50}{\includegraphics{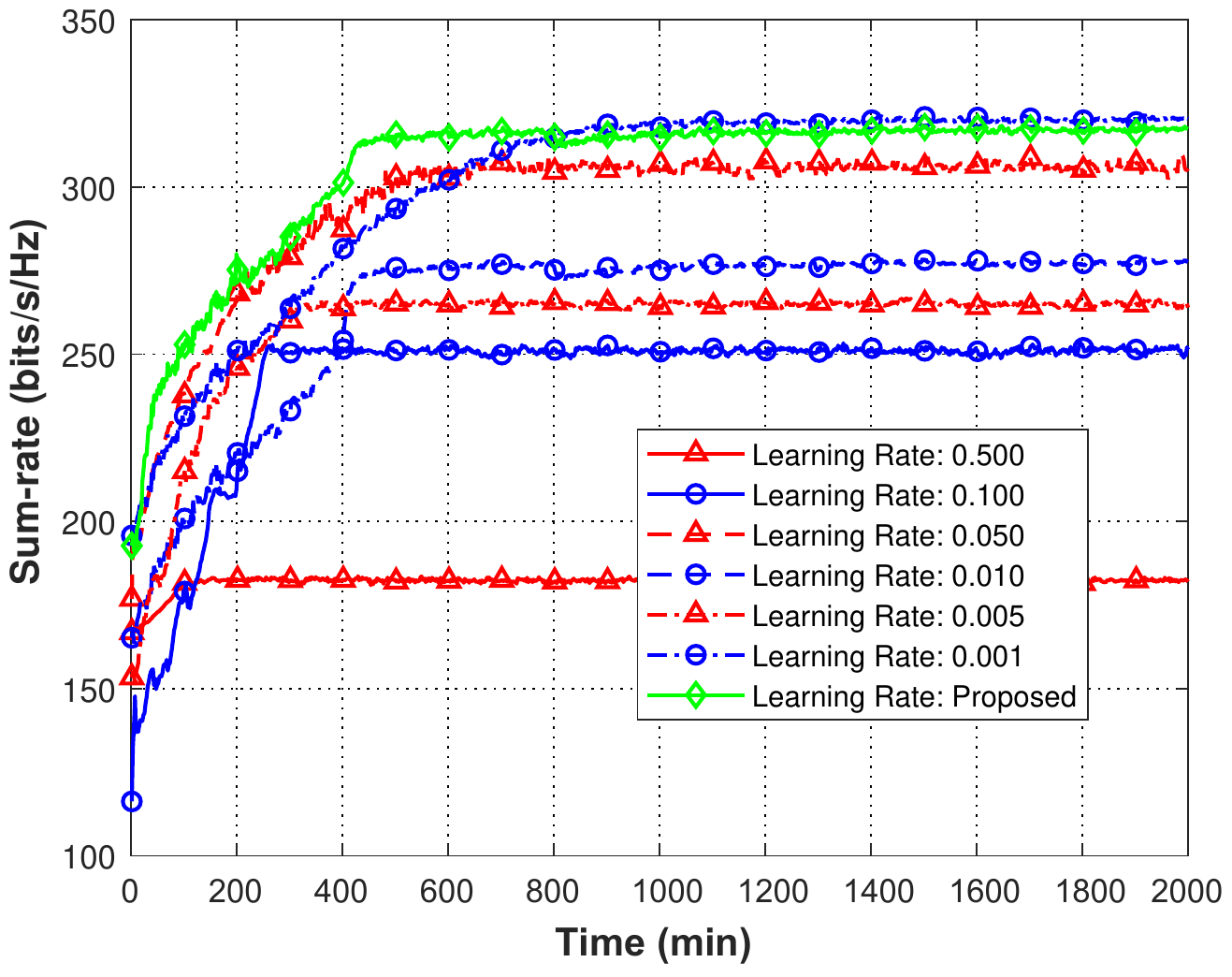}} }	
    \caption{Convergence performance for different numbers of batch size and different choices of learning rate $(N_{t}=128, K=30)$.}
	\label{BatchRate}
	\end{figure}

\begin{figure}[!t]
	\centering
	\subfloat[$K=10$.]{\centering \scalebox{0.50}{\includegraphics{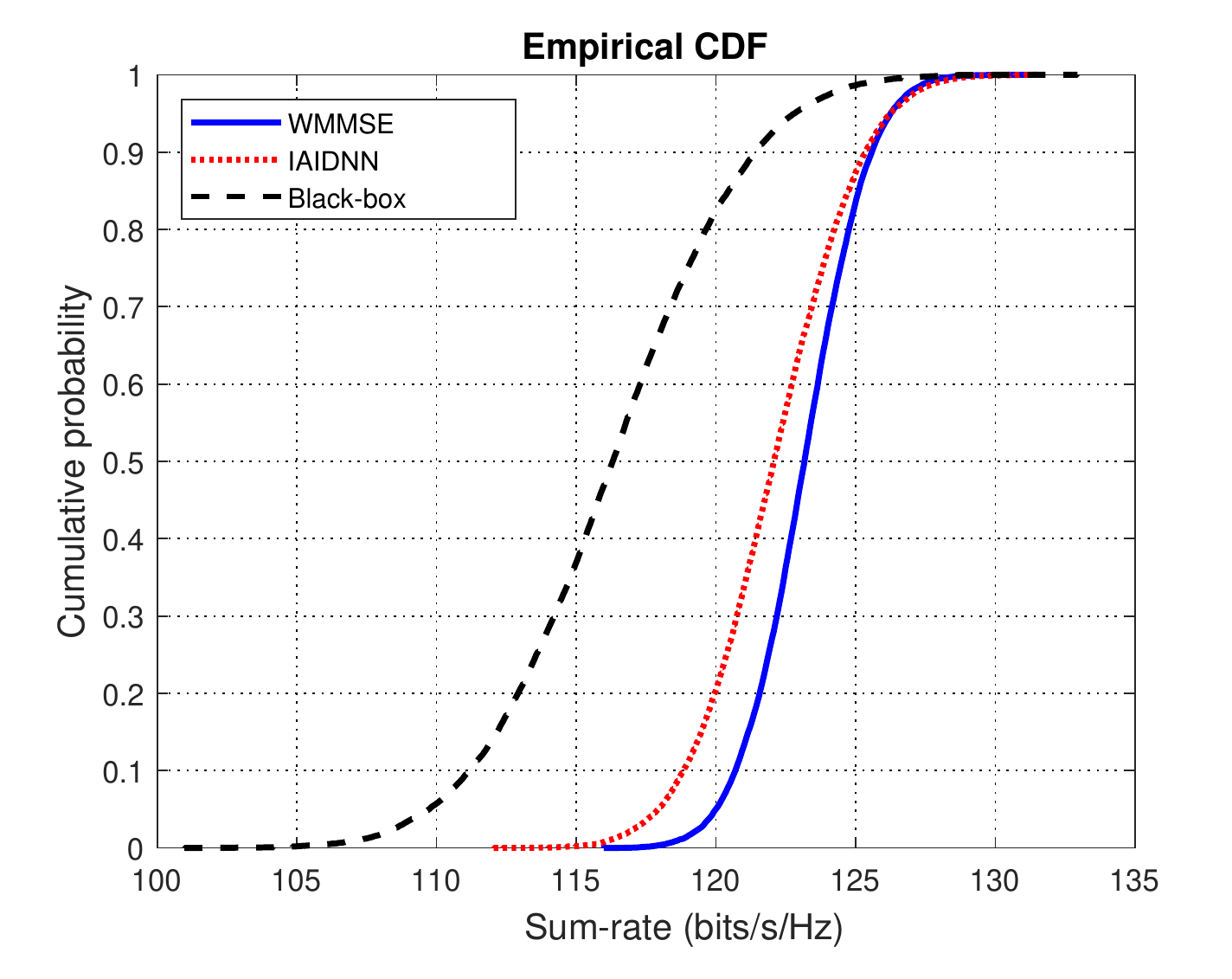}} }
	\subfloat[$K=30$.]{\centering \scalebox{0.50}{\includegraphics{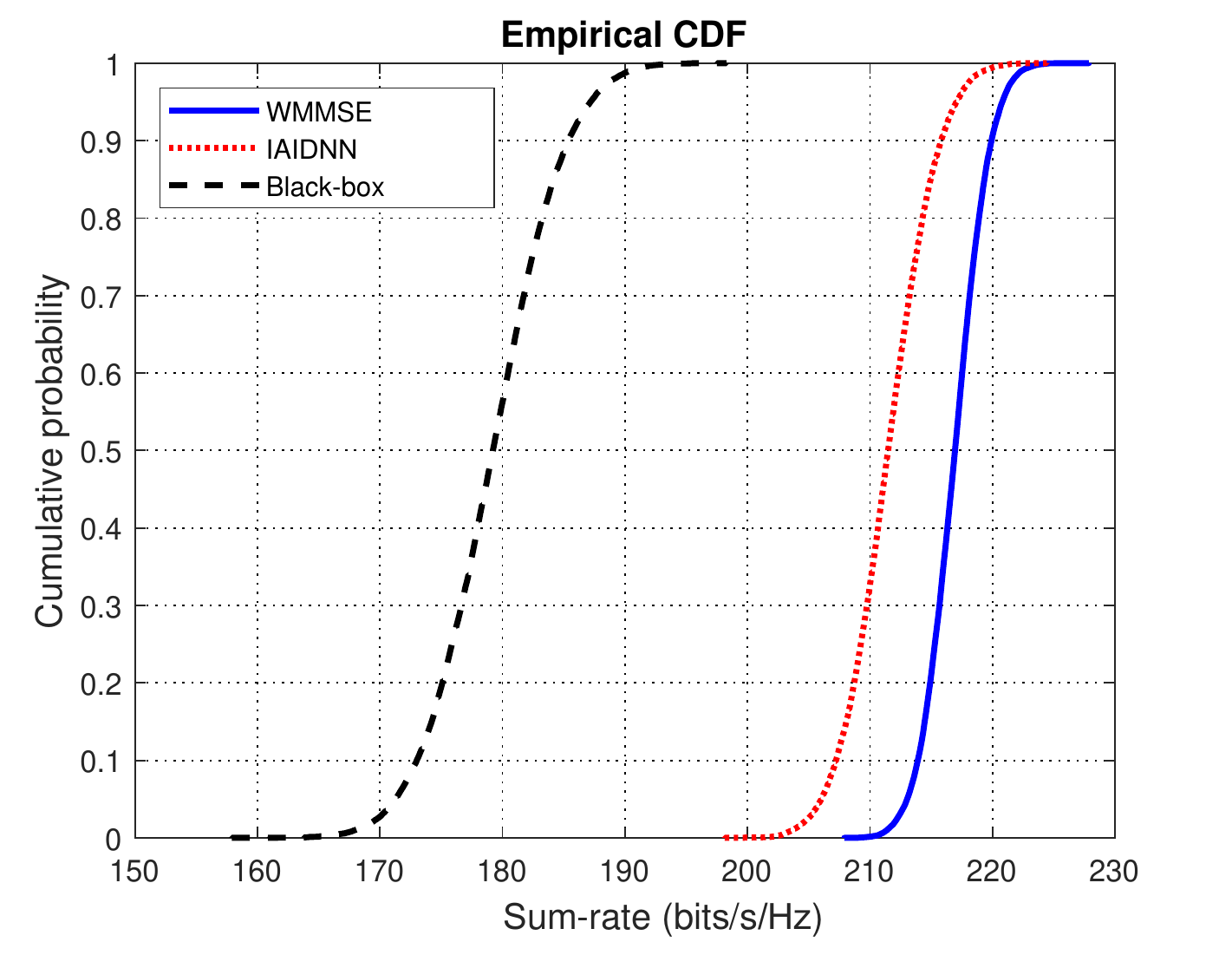}} }	
	\caption{The CDF that describes the sum-rate achieved by different schemes for $N_{t}=64$.}
	\label{CDFNt64}
\end{figure}

\begin{figure}[!t]
	\centering
	\subfloat[$K=20$.]{\centering \scalebox{0.50}{\includegraphics{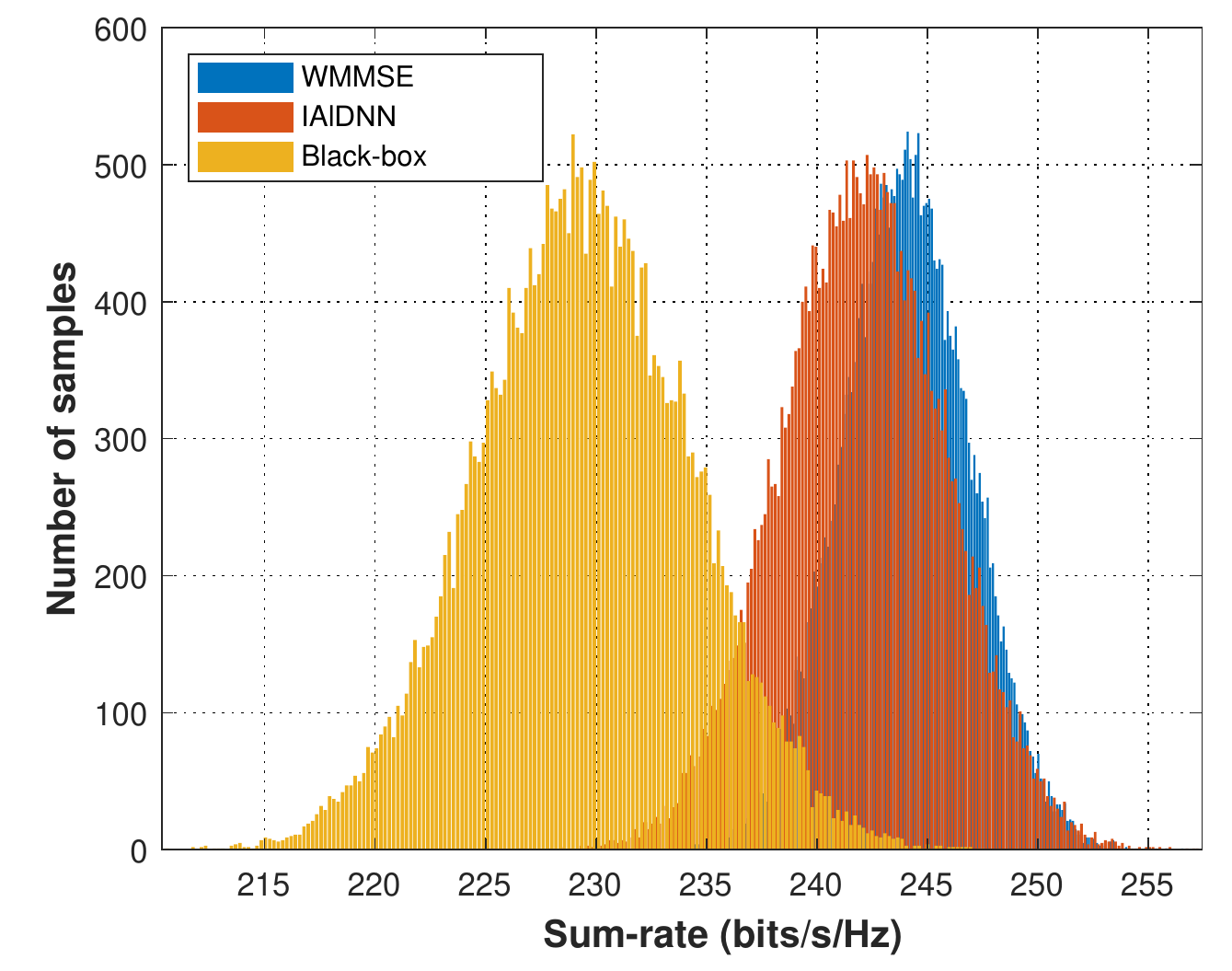}} }
	\subfloat[$K=60$.]{\centering \scalebox{0.50}{\includegraphics{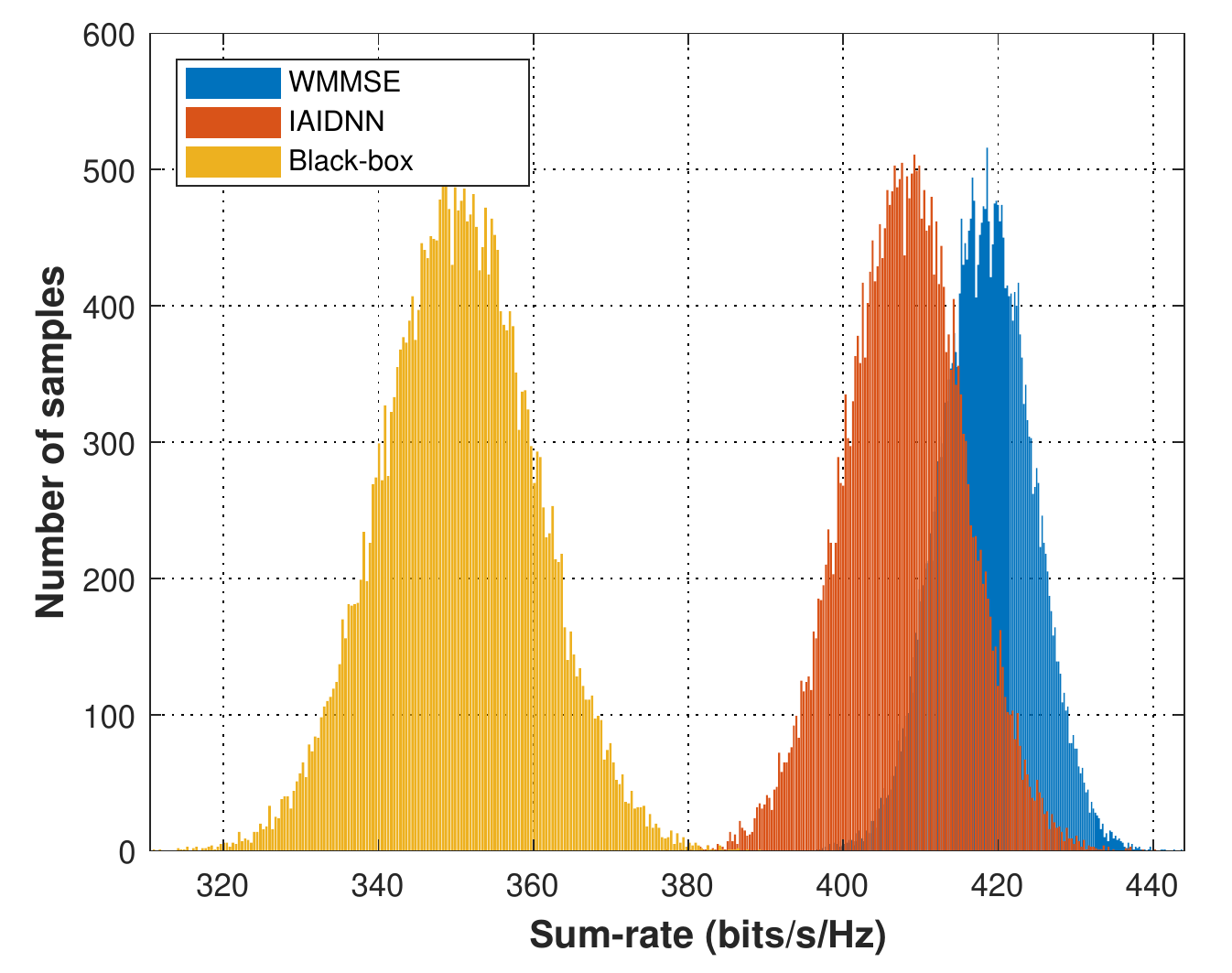}} }	
	\caption{Distributions of different schemes for $N_{t}=128$.}
	\label{DisNt128}
\end{figure}

Fig. \ref{BatchRate} presents the impact of the batch size and learning rate on the convergence performance. From Fig. \ref{BatchRate} (a), a larger batch size leads to slower but more stable convergence performance, while the achieved sum-rate increases with the decrease of batch size. It is mainly because the randomness of the gradient becomes larger with the decrease of batch size, which increases the possibility for the IAIDNN to bypass the saddle point and find the globally optimal solution. From Fig. \ref{BatchRate} (b), we can see that a smaller learning rate achieves better sum-rate performance, while a larger learning rate leads to faster convergence performance, and the proposed learning rate scheme shows a good balance between the convergence speed and the sum-rate performance.

Fig. \ref{CDFNt64} shows the cumulative distribution function (CDF) that describes the sum-rate performance achieved by different investigated schemes for $N_{t}=64$, where $50,000$ testing data samples of $\{\mathbf{H}_{k}, \forall k\}$ are generated. We can see that when the number of users $K$ is small, e.g., $K=10$, the proposed IAIDNN achieves $99.82\%$ sum-rate performance of the iterative WMMSE algorithm on average, while that of the black-box based CNN is $92.83\%$. The gaps among the three schemes increase with $K$. Moreover, the variance of the results achieved by the iterative WMMSE algorithm is the smallest, while that of the black-box based CNN is the largest among these schemes.

Fig. \ref{DisNt128} presents the distribution of the sum-rate over the entire test data set for $N_{t}=128$. It is observed that the proposed IAIDNN provides a good approximation of the entire rate profile generated by the iterative WMMSE algorithm, whose approximation is better than that of the black-box based CNN. The approximation becomes worse when $K$ increases, especially when the value of $K\times N_{r}$ approaches $N_{t}$.

\begin{table*}[t]
	\caption{The sum-rate performance versus SNR $(N_{t}=64, K=30)$.}
	\begin{center}
		\centering
		\begin{tabular}{cccccccc}
            \rowcolor{mygray}
			\thickhline
			\centering
			\textbf{SNR (dB)}    & $0$ & $5$ & $10$  & $15$ & $20$ & $25$ & $30$    \\
			\textbf{WMMSE (bits/s/Hz)}  & $45.86$ & $85.52$ & $118.32$  & $164.77$ & $216.92$ & $266.55$ & $317.18$  \\
            \rowcolor{mygray}
			\textbf{IAIDNN}  & $97.06\%$ & $ 97.11\%$ & $ 97.25\%$ & $97.32\%$ & $97.58\% $ & $97.79\%$ & $ 98.02\%$ \\
            \textbf{Black-box} & $80.32\%$ & $80.54\%$ & $80.87\%$ & $81.32\%$  & $81.93\% $ & $82.65\%$ & $83.72\% $  \\
			\thickhline
		\end{tabular}
	\end{center}
	\label{SNR}
\end{table*}

Table \ref{SNR} presents the performance versus SNR in the scenario of $(N_{t}=64, K=30)$. The sum-rate performance achieved by the IAIDNN and black-box based CNN is slightly improved with the increase of SNR.
It is mainly because the feasible region of the problem under investigation expands when SNR increases, then the network tends to find a better solution.

\begin{table*}[t]
	\caption{The sum-rate performance versus \# of training samples $(N_{t}=128, K=40)$.}
	\begin{center}
		\centering
		\begin{tabular}{ccccccccc}
            \rowcolor{mygray}
            \thickhline
			\centering
			\textbf{\# of training samples}   & $5000$  & $10000$ & $15000$  & $20000$ & $25000$ & $30000$ & $35000$ &$40000$ \\
			\textbf{Black-box}  & $78.93\%$ & $83.14\%$ & $86.59\%$  & $88.36\%$ & $89.66\%$ & $89.93\%$ & $ 90.15\%$ & $90.15\%$  \\  \hline
            \rowcolor{mygray}
            \textbf{\# of training samples}   & $100$  & $200$ & $300$  & $400$ & $500$ & $600$ & $700$  & $800$  \\
			\textbf{IAIDNN}  & $93.03\%$ & $95.89\%$ & $97.31\%$  & $98.52\%$ & $98.94\%$ & $99.03\%$ & $99.26\%$ & $99.26\%$  \\
			\thickhline
		\end{tabular}
	\end{center}
	\label{Sample}
\end{table*}

Table \ref{Sample} shows the sum-rate performance versus the number of training data samples for the case of $(N_{t}=128, K=40)$. It is obvious that the proposed IAIDNN needs much fewer training data samples than the conventional black-box based CNN since it makes use of the structure of the classic iterative WMMSE algorithm. This advantage is important in a realistic industrial application due to the challenge of obtaining training data samples and the high cost of implementing channel estimation.

\begin{table*}[t]
	\caption{The sum-rate performance versus \# of layers $(N_{t}=64, K=30)$.}
	\begin{center}
		\centering
		\begin{tabular}{ccccccccc}
            \rowcolor{mygray}
            \thickhline
			\centering
			\textbf{\# of layers}   & $3$  & $4$ & $5$  & $6$ & $7$ & $8$ & $9$ & $10$  \\
			\textbf{Sum-rate performance}  & $91.58\%$ & $93.35\%$ & $95.56\%$  & $96.61\%$ & $97.58\%$ & $97.61\%$ & $97.34\%$ & $96.93\%$ \\
            \rowcolor{mygray}
            \textbf{CPU time of training (min) }  & $265.63$ & $279.35$ & $289.02$  & $296.97$ & $301.61$ & $306.16$ & $310.91$ & $313.56$ \\
			\thickhline
		\end{tabular}
	\end{center}
	\label{Layer}
\end{table*}

Table \ref{Layer} shows that with the increase of layer $L$, the sum-rate performance improves first, and then decreases. It is because when $L$ is small, the degree of freedom of the IAIDNN is small, which leads to its unsatisfactory learning ability with small number of trainable parameters. Thus, the performance improves when $L$ increases. However, the numerical error of the gradients increases with $L$ due to a series of operations of matrix inversion and multiplication. When $L$ is relatively large, e.g., $L=8$, the learning ability of the network is limited by the numerical error, which leads to the degrade of the sum-rate performance.
Moreover, the training time also increases with $L$, and $L=7$ is the optimal choice since it achieves a good balance between the performance and training time.

\begin{figure}[t]
	\begin{centering}
		\includegraphics[width=0.5\textwidth]{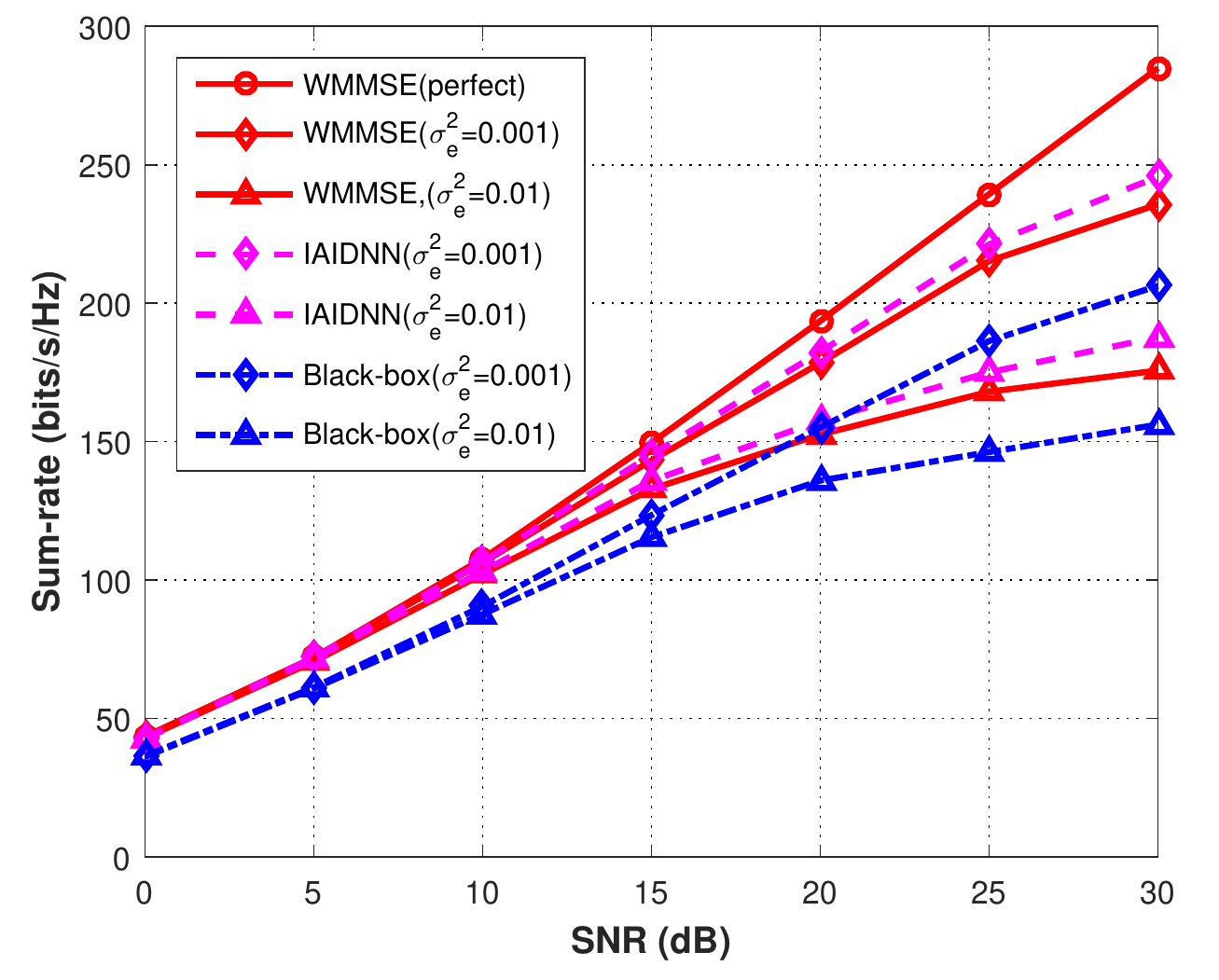}
		\par\end{centering}
	\caption{The sum-rate performance in the scenario of imperfect CSI $(N_{t}=64, K=20)$.}
	\label{CSI}
\end{figure}

Fig. \ref{CSI} illustrates the sum-rate performance of the analyzed algorithms in the scenario of imperfect CSI. From the results, the performance degrades with the increase of the CSI error variance $\sigma_e^2$. The proposed IAIDNN provides the best performance, followed by the iterative WMMSE algorithm and the black-box network, which shows the ability of the proposed IAIDNN to handle channel uncertainties. Since the IAIDNN aims at maximizing the average sum-rate, it has better robustness compared to the iterative WMMSE algorithm.

\subsection{Complexity Comparison}
\begin{table*}[t]
	\caption{The CPU running time of the analyzed schemes.}
	\begin{center}
		\centering
		\begin{tabular}{c"cc"ccc}
        \thickhline
            \multirow{2}{*}{\begin{tabular}[c]{@{}c@{}} \textbf{\# of transmit antennas} \\ \textbf{and users $(N_{t},K)$}\end{tabular}} & \multicolumn{2}{c"}{\textbf{CPU time of training stage (min)}} & \multicolumn{3}{c}{\textbf{CPU time of testing stage (s)}}  \\  
              & \textbf{IAIDNN}    & \textbf{Black-box}   & \textbf{IAIDNN} & \textbf{Black-box} & \textbf{WMMSE}      \\ \hline
           \rowcolor{mygray}
           \textbf{(8,4)}   &8.71     &11.12   &0.01   &0.01   &0.08       \\
           \textbf{(16,8)}  &21.52    &31.55   &0.01   &0.01   &0.22       \\
           \rowcolor{mygray}
           \textbf{(32,8)}  &28.65    &60.18   &0.02   &0.03   &0.51       \\
           \textbf{(32,16)} &70.35    &130.54  &0.03   &0.04   &1.05       \\
           \rowcolor{mygray}
           \textbf{(64,15)} &102.23   &153.66  &0.04   &0.05   &1.46       \\
           \textbf{(64,30)} &301.61   &467.12  &0.11   &0.13   &4.22       \\
           \rowcolor{mygray}
           \textbf{(128,30)}&514.56   &1439.43 &0.13   &0.16   &5.14       \\
           \textbf{(128,60)}&1242.23  &4184.02 &0.32   &0.39   &29.68       \\
           \rowcolor{mygray}
           \textbf{(256,30)}&1056.59  &3758.51 &0.61   &0.71   &32.31       \\
           \textbf{(256,60)}&3126.12  &9986.37 &0.83   &0.98   &38.56       \\
           \rowcolor{mygray}
           \textbf{(256,120)}&9806.85 &$-$ &2.94   &3.36   &291.01       \\ \thickhline
		\end{tabular}
	\end{center}
	\label{Time}
\end{table*}

Table \ref{Time} compares the computational complexity, i.e., the CPU time of the training stage and testing stage, for different schemes in various scenarios. It is obvious that the CPU time of the training stage and testing stage both increases with the number of transmit antennas $N_{t}$ and the number of users $K$, while the CPU time of the training stage grows much more quickly than that of the testing stage since the training stage has more operations of matrix multiplication and inversion. Moreover, the proposed IAIDNN has much shorter training time and converges faster than the black-box based CNN. It is because the loss function \eqref{objlossfunc} in the unsupervised learning is complicated, the black-box based CNN applying the ``tensorflow" for calculating the gradients in the BP computation is not efficient.
In the proposed IAIDNN, we derive the closed-form gradients in the BP computation efficiently. The gap of the CPU time between the IAIDNN and the black-box based CNN in the training stage becomes larger when $N_{t}$ and $K$ increase. In the testing stage, we can see that the IAIDNN requires shorter CPU time than that of the iterative WMMSE algorithm, where the superiority of our proposed algorithm lies in. In a large-scale MU-MIMO scenario, i.e., $N_{t}=128$ or $256$, this superiority is much significant, which
makes it possible that iterative algorithms can be widely used in practical engineering.

\subsection{Generalization Ability}
\begin{table*}[t]
	\caption{The generalization ability of the IAIDNN.}
	\begin{center}
		\centering
		\begin{tabular}{c"ccccccc}
			\thickhline
			\centering
			\diagbox{ \textbf{ \# of antennas} $\mathbf{N_{t}}$}{ \textbf{ \# of users }$\mathbf{K}$}  & $\mathbf{70}$  & $\mathbf{60}$  & $\mathbf{50}$ & $\mathbf{40}$  & $\mathbf{30}$ & $\mathbf{20}$ &
                                               $\mathbf{10}$    \\   \hline
            \rowcolor{mygray}
			\textbf{256}  & $98.07\%$ & $ 98.25\%$ & $98.36\%$ & $98.51\%$  & $98.84\%$ & $99.01\%$ & $99.12\%$    \\
            \textbf{128}  & $-$ & $95.12\%$ & $95.97\%$ & $96.67\%$  & $96.93\%$ & $97.21\%$ & $97.43\%$  \\
            \rowcolor{mygray}
            \textbf{64}   & $-$ & $-$ & $-$ & $-$ & $94.69\%$ & $96.82\%$ & $97.75\%$  \\
			\thickhline
		\end{tabular}
	\end{center}
	\label{Scalab}
\end{table*}

Table \ref{Scalab} shows the generalization ability of the proposed IAIDNN. We train a network with $N_{t}=256$ transmit antennas and $K=80$ users, and then apply this network to test the sum-rate performance of the scenarios with smaller $N_{t}$ and $K$ in Table \ref{Scalab}. By comparing these results to those shown in Table \ref{Nt64}, \ref{Nt128}, and \ref{Nt256}, we can see that the performance loss of applying this large-scale network to test the scenarios with smaller number of users and the same number of transmit antennas is around $1\%$, while that of the scenarios with smaller number of users and transmit antennas is around $3\%$.
Thus, the generalization ability of the proposed IAIDNN is satisfactory.

%

\subsection{Improvement of the IAIDNN for Fully Loaded Systems}
The proposed IAIDNN achieves good sum-rate performance in the case of $K\times N_{r}<N_{t}$, which approaches that of the iterative WMMSE algorithm. However, its sum-rate performance degrades in a fully loaded system, i.e., $K\times N_{r}=N_{t}$. In the following, we modify the structure of the network to improve its performance in this case. 
We introduce the matrix inversion operation $\mathbf{A}^{-1}$ and include more parameters, i.e., $\mathbf{P}$ to increase the degree of freedom.
Then, the structure $\mathbf{A}^{-1}\mathbf{X} + \mathbf{P}\mathbf{A}\mathbf{Y}+ \mathbf{Z}$ is applied to replace $\mathbf{A}^{+}\mathbf{X} + \mathbf{A}\mathbf{Y} + \mathbf{Z}$ in \eqref{networkUWV}, where $\mathbf{X}$, $\mathbf{P}$, $\mathbf{Y}$, and $\mathbf{Z}$ are introduced trainable parameters.
The training stage and the testing stage are the same as the IAIDNN proposed in Section \ref{AlgoInduce}.

\begin{table*}[t]
	\caption{The sum-rate performance of the improved IAIDNN in a fully loaded system.}
	\begin{center}
		\centering
		\begin{tabular}{cccccc}
            \rowcolor{mygray}
			\thickhline
			\centering
			\textbf{Scenario $(N_{t},K)$}    & $(8,4)$ & $(16,8)$ & $(32,16)$  & $(64,32)$ & $(128,64)$     \\
			\textbf{WMMSE in $7$ iterations}  & $79.19\%$ & $80.63\%$ & $82.72\%$  & $83.05\%$ & $83.52\%$   \\
            \rowcolor{mygray}
            \textbf{IAIDNN}  & $91.35\%$ & $92.13\%$ & $92.63\%$ & $92.82\%$ & $92.95\%$  \\
            \textbf{Improved IAIDNN}  & $95.86\%$ & $ 96.02\%$ & $ 96.85\%$ & $96.93\%$ & $97.08\%$  \\
            \rowcolor{mygray}
            \textbf{Training time (min)}  & $10.65$ & $27.52$ & $86.37$ & $372.58$ & $2112.31$  \\
			\thickhline
		\end{tabular}
	\end{center}
	\label{Improvement1}
\end{table*}

Table \ref{Improvement1} shows that the improvement of the proposed IAIDNN significantly increases the sum-rate performance and even outperforms the classic iterative WMMSE algorithm in $7$ iterations. However, the improved IAIDNN has more trainable parameters and requires a few matrix inversion operations, which causes slightly increased computational complexity and training time.

Based on the simulation results and discussion presented above, we summarize some features of the analyzed schemes in Table \ref{feature}.

\begin{table*}[t]
	\caption{Features of the analyzed schemes.}
	\begin{center}
		\centering
		\begin{tabular}{ccccccc}
			\hline
			\centering
			\textbf{Algorithms}   & Performance & Efficiency &   
			Parameter Dimension & Robustness & Interpretability & Generalizability  \\
			\hline
			\textbf{WMMSE} & High & Low & Low & Low & High  & Low  \\
			\hline
			\textbf{IAIDNN} & High & High & Middle & High & Middle  & Middle  \\
			\hline
			\textbf{Black-box} & Low & High & High & Middle & Low  & Middle \\
			\hline
		\end{tabular}
	\end{center}
	\label{feature}
\end{table*}

\section{Conclusion}
\label{Conclusion}
In this work, we proposed a novel deep-unfolding based framework, where a general form of IAIDNN in matrix form is developed. 
To design the precoding in MU-MIMO systems, we developed an IAIDNN based on the structure of the classic WMMSE iterative algorithm. Specifically, the iterative WMMSE algorithm is unfolded into a layer-wise structure in the IAIDNN, where a number of trainable parameters are introduced to replace the high-complexity operations.
To train the network, a GCR of the IAIDNN has been proposed to depict the recurrence relation between two adjacent layers in BP. 
Simulation results showed that the proposed IAIDNN can be trained to efficiently achieve the performance of the iterative WMMSE algorithm with reduced computational complexity.
Thus, we can conclude that IAIDNNs can be applied as surrogates of the iterative optimization algorithms in real-time systems. 
The future work could generalize our proposed IAIDNN framework in matrix form to other challenging communication applications,
such as the robust precoding design in the presence of CSI errors, the precoding design in a multicell system, and the problem with discrete variables.

\begin{appendices}	
\section{ Proof for Theorem \ref{theorem Chain rule} }
\label{Appendix 0}

Firstly, we introduce the following Theorem \ref{theorem matrix diff} to compute the gradient of a matrix variable. Then, the properties of matrix differential are presented in Lemma \ref{Properties}.
	
\begin{theorem} \label{theorem matrix diff}
	If the differential of a function $f$ with matrix variable $\mathbf{X}$ has the following form
	\begin{equation}
	df = \textrm{Tr}(\mathbf{A}d\mathbf{X}^H), 
	\end{equation}
	then, the partial derivative of $f$ with respect to $\mathbf{X}^{*}$ is \cite{MatrixDiff}:
	\begin{equation}
	\frac{\partial f}{\partial \mathbf{X}^*} = \mathbf{A}, 
	\end{equation}
	where $f:\mathbb{C}^{m\times n}\mapsto \mathbb{R}$ is a function with respect to variable $\mathbf{X}\in \mathbb{C}^{m\times n}$.
\end{theorem}

\begin{lemma} \label{Properties}
	The properties of matrix differential \cite{MatrixDiff}:
	\begin{equation}  \label{diffpro}
	\begin{aligned}
	& d\textrm{Tr}(\mathbf{X})=\textrm{Tr}(d\mathbf{X}), \quad \quad \quad \quad \quad d\textrm{Tr}(\mathbf{X}\mathbf{X}^{H})=\textrm{Tr}(\mathbf{X}d\mathbf{X}^{H}+\mathbf{X}^{H}d\mathbf{X}),  \\
	& d\log \det(\mathbf{X})=\textrm{Tr}(\mathbf{X}^{-1}d\mathbf{X}), \quad  d\textrm{Tr}(\mathbf{A} \mathbf{X}^{-1})=-\textrm{Tr}(\mathbf{X}^{-1}\mathbf{A} \mathbf{X}^{-1} d\mathbf{X}), \\
	& d(\mathbf{X}+\mathbf{Y})=d\mathbf{X}+d\mathbf{Y}, \quad \quad \quad \!\! d(\mathbf{X}\mathbf{Y})=(d\mathbf{X})\mathbf{Y}+\mathbf{X}(d\mathbf{Y}),
	\end{aligned}
	\end{equation}
	where $\mathbf{A}$ is a constant matrix, $\mathbf{X}$ and $\mathbf{Y}$ are matrix variables.
\end{lemma}

Recall the properties of the trace of the matrix,
\begin{equation} \label{tracepro}
\begin{aligned}
& & \textrm{Tr}(\mathbf{A}\mathbf{B})=\textrm{Tr}(\mathbf{B}\mathbf{A}),  \quad
\textrm{Tr}\big( \mathbf{A}^{T}(\mathbf{B}\circ \mathbf{C}) \big)=\textrm{Tr}\big( (\mathbf{A}^{T}\circ \mathbf{B}^{T})\mathbf{C} \big).
\end{aligned}
\end{equation}
Based on Theorem \ref{theorem matrix diff}, Lemma \ref{Properties}, and \eqref{tracepro}, the GCR in matrix form in Theorem \ref{theorem Chain rule} is obtained.

\section{The Gradient of \eqref{objlossfunc0} with Respect to $\mathbf{U}_k^{L}$ and $\mathbf{W}_k^{L}$ }
\label{Appendix A}

Based on Theorem \ref{theorem matrix diff} and Lemma \ref{Properties},
the gradient of the objective function \eqref{objlossfunc0} with respect to $\mathbf{W}_k^{L}$ in the last layer is presented below. For clarity, we omit the index of layer $l$, where the variables here are all from the last layer, i.e., $l=L$.
\begin{equation} \label{gradientW}
\begin{aligned}
\frac{\partial f}{\partial \mathbf{W}_{k}^{L}} &= - \sum\limits_{m\neq k} \mathbf{U}_{k}^{H}\mathbf{H}_{k}\tilde{\mathbf{C}}^{-H}\mathbf{H}_{m}^{H}\tilde{\mathbf{E}}_{m}\mathbf{H}_{m}\mathbf{V}_{k}
- \sum\limits_{m=1}^K \textrm{Tr}(  \tilde{\mathbf{A}}_{m}^{-1}\mathbf{H}_{m}\tilde{\mathbf{D}}_{m}\tilde{\mathbf{C}}^{-H}\mathbf{H}_{m}^{H}\tilde{\mathbf{B}}_{m}^{-1})
\frac{\sigma_{k}^{2}}{P_{T}}\mathbf{U}_{k}^{H}\mathbf{U}_{k}     \\
& \quad  - \sum\limits_{m=1}^K \frac{\sigma_{k}^{2}}{P_{T}}\textrm{Tr}(\tilde{\mathbf{E}}_{m})\mathbf{U}_{k}^{H}\mathbf{H}_{k}\tilde{\mathbf{C}}^{-H}\mathbf{V}_{k}
-\sum\limits_{m=1}^K \mathbf{U}_{k}^{H}\mathbf{H}_{k}\tilde{\mathbf{C}}^{-H}\mathbf{H}_{m}^{H}\tilde{\mathbf{B}}_{m}^{-1}
\tilde{\mathbf{A}}_{m}^{-1}\mathbf{H}_{m}\tilde{\mathbf{D}}_{m}\mathbf{H}_{k}^{H}\mathbf{U}_{k}      \\
& \quad  + \mathbf{U}_{k}^{H}\mathbf{H}_{k}\tilde{\mathbf{C}}^{-H}\mathbf{H}_{k}^{H}\tilde{\mathbf{B}}_{k}^{-1}\tilde{\mathbf{A}}_{k}^{-1}\mathbf{H}_{k}\mathbf{V}_{k}
+ \sum\limits_{n=1}^K \sum\limits_{m\neq k} \mathbf{U}_{k}^{H}\mathbf{H}_{k}\tilde{\mathbf{C}}^{-H}\mathbf{H}_{n}^{H}\tilde{\mathbf{E}}_{n}\mathbf{H}_{n}\tilde{\mathbf{D}}_{m}\mathbf{H}_{k}^{H}\mathbf{U}_{k} \\
& \quad  + \! \sum\limits_{n=1}^K \! \sum\limits_{m=1}^K \! \frac{\sigma_{n}^{2}\sigma_{k}^{2}}{P_{T}^{2}}\textrm{Tr}( \tilde{\mathbf{E}}_{m} \! )\textrm{Tr}( \tilde{\mathbf{D}}_{n}\tilde{\mathbf{C}}^{-H} \! )\mathbf{U}_{k}^{H}\mathbf{U}_{k}
\!+\! \sum\limits_{n=1}^K \sum\limits_{m\neq k} \textrm{Tr}( \tilde{\mathbf{E}}_{n}\mathbf{H}_{n}\tilde{\mathbf{D}}_{m}\tilde{\mathbf{C}}^{-H}\mathbf{H}_{n}^{H} \! ) \frac{\sigma_{k}^{2}}{P_{T}}\mathbf{U}_{k}^{H}\mathbf{U}_{k} \\
& \quad  + \! \sum\limits_{n=1}^K \sum\limits_{m=1}^K \frac{\sigma_{n}^{2}}{P_{T}}\textrm{Tr}( \tilde{\mathbf{E}}_{m})
\mathbf{U}_{k}^{H}\mathbf{H}_{k}\tilde{\mathbf{C}}^{-H}\tilde{\mathbf{D}}_{n}\mathbf{H}_{k}^{H}\mathbf{U}_{k},
\end{aligned}
\end{equation}
where
\begin{subequations}
\begin{eqnarray}
& & \tilde{\mathbf{A}}_{k}\triangleq  \mathbf{H}_{k}\mathbf{V}_{k}\mathbf{V}_{k}^{H}\mathbf{H}_{k}^{H}\tilde{\mathbf{B}}_{k}^{-1} ,  \notag \\
& & \tilde{\mathbf{B}}_{k}\triangleq  \sum\limits_{m\neq k} \mathbf{H}_{k}\mathbf{V}_{m}\mathbf{V}_{m}^{H}\mathbf{H}_{k}^{H}
+ \frac{\sigma_{k}^{2}}{P_{T}} \sum\limits_{k=1}^K \textrm{Tr}(\mathbf{V}_{k}\mathbf{V}_{k}^{H})\mathbf{I} ,     \notag \\
& & \tilde{\mathbf{C}}\triangleq  \sum\limits_{k=1}^K \frac{\sigma_{k}^{2}}{P_{T}} \textrm{Tr}(\mathbf{U}_{k}\mathbf{W}_{k}\mathbf{U}_{k}^{H})\mathbf{I} + \sum\limits_{m=1}^K \mathbf{H}_{m}^{H}\mathbf{U}_{m}\mathbf{W}_{m}\mathbf{U}_{m}^{H}\mathbf{H}_{m}   ,    \notag  \\
& & \tilde{\mathbf{D}}_{k}\triangleq  \mathbf{V}_{k}\mathbf{W}_{k}^{H}\mathbf{U}_{k}^{H}\mathbf{H}_{k} \tilde{\mathbf{C}}^{-H} ,   \quad
\tilde{\mathbf{E}}_{k}\triangleq  \tilde{\mathbf{B}}_{k}^{-1}\tilde{\mathbf{A}}_{k}^{-1}\mathbf{H}_{k}\mathbf{V}_{k}\mathbf{V}_{k}^{H}\mathbf{H}_{k}^{H}\tilde{\mathbf{B}}_{k}^{-1} . \notag
\end{eqnarray}
\end{subequations}
The gradient of objective function with respect to $\mathbf{U}_k^{L}$ in the last layer can be computed similarly.

\section{ Details of Calculating the Gradients $\{ \mathbf{G}_{n}^{u,l}, \mathbf{G}_{n}^{w,l}, \mathbf{G}_{n}^{v,l}, \forall l,n \}$ }
\label{Appendix B}

The total gradient with respect to all the matrix variables at the $(l+1)$-th layer is given by
\begin{equation}
\sum\limits_{k=1}^K \textrm{Tr}\bigg( \mathbf{G}_{k}^{u,l+1}d\mathbf{U}_{k}^{l+1} + \mathbf{G}_{k}^{w,l+1}d\mathbf{W}_{k}^{l+1} + \mathbf{G}_{k}^{v,l+1}d\mathbf{V}_{k}^{l+1} \bigg). \label{totaldiff}
\end{equation}
We take $\mathbf{G}_{n}^{w,l}$ as an example, $\mathbf{G}_{n}^{u,l}$ and $\mathbf{G}_{n}^{v,l}$ can be obtained similarly. In order to calculate $\mathbf{G}_{n}^{w,l}$, we firstly substitute \eqref{networkUWV} into \eqref{totaldiff}, and the corresponding details are shown in Appendix \ref{Appendix C}. Then, we retain the terms with $d\mathbf{W}_{k}^{l}$ and have the following results based on Theorem \ref{theorem Chain rule} and Corollary \ref{Case},
\begin{equation}  \label{dW}
\begin{aligned}
& \textrm{Tr}\bigg\{  \mathbf{G}_{n}^{v,l+1} \bigg( -(\mathbf{B}^{l+1})^{+}\circ (\mathbf{B}^{l+1})^{+}\circ d(\mathbf{B}^{l+1}) \bigg) \mathbf{X}_{n}^{v,l+1}\mathbf{H}_{n}^{H} \mathbf{U}_{n}^{l+1} \mathbf{W}_{n}^{l+1}    \\
& \quad \quad + \mathbf{G}_{n}^{v,l+1} d(\mathbf{B}^{l+1}) \mathbf{Y}_{n}^{v,l+1}\mathbf{H}_{n}^{H} \mathbf{U}_{n}^{l+1}\mathbf{W}_{n}^{l+1}
+ \mathbf{G}_{n}^{v,l+1} \mathbf{M}_{n}^{v,l+1} \mathbf{U}_{n}^{l+1} d\mathbf{W}_{n}^{l+1} \bigg\} \\
&\overset{\eqref{Chain rule}}{=} \textrm{Tr}\bigg\{ \sum\limits_{k=1}^K \textrm{Tr}( \mathbf{J}_{k}^{v,l+1}+\mathbf{L}_{k}^{v,l+1} )\frac{\sigma^{2}_{k}}{P_{T}}(\mathbf{U}_{n}^{l+1})^{H}\mathbf{U}_{n}^{l+1}d\mathbf{W}_{n}^{l+1} \\
& \quad \quad \quad + \sum\limits_{k=1}^K (\mathbf{U}_{n}^{l+1})^{H}\mathbf{H}_{n}( \mathbf{J}_{k}^{v,l+1}+\mathbf{L}_{k}^{v,l+1} )\mathbf{H}_{n}^{H}\mathbf{U}_{n}^{l+1}d\mathbf{W}_{n}^{l+1}
+ \mathbf{G}_{n}^{v,l+1}\mathbf{M}_{n}^{v,l+1}\mathbf{U}_{n}^{l+1} d\mathbf{W}_{n}^{l+1} \bigg\},
\end{aligned}
\end{equation}
where
\begin{equation}
\begin{aligned}
& d(\mathbf{B}^{l+1}) = \sum\limits_{k=1}^K \frac{\sigma^{2}_{k}}{P_{T}} \textrm{Tr}(\omega_k
\mathbf{U}^{l+1}_{k}d\mathbf{W}^{l+1}_{k}(\mathbf{U}^{l+1}_{k})^{H})\mathbf{I}
+ \sum\limits_{m=1}^K \omega_m \mathbf{H}_{m}^{H}\mathbf{U}^{l+1}_{m}d\mathbf{W}^{l+1}_{m}(\mathbf{U}^{l+1}_{m})^{H}\mathbf{H}_{m} ,  \\
& \mathbf{J}_{k}^{v,l+1} \triangleq \bigg( \mathbf{X}_{k}^{v,l+1}\mathbf{H}_{k}^{H}\mathbf{U}_{k}^{l+1}\mathbf{W}_{k}^{l+1}\mathbf{G}_{k}^{v,l+1} \bigg)\circ \bigg( -(\mathbf{B}^{l+1})^{+}\circ (\mathbf{B}^{l+1})^{+} \bigg)^{T} ,  \\
& \mathbf{L}_{k}^{v,l+1} \triangleq \mathbf{Y}_{k}^{v,l+1}\mathbf{H}_{k}^{H}\mathbf{U}_{k}^{l+1}\mathbf{W}_{k}^{l+1}\mathbf{G}_{k}^{v,l+1} ,  \\
& \mathbf{M}_{k}^{v,l+1} \triangleq  \bigg( (\mathbf{B}^{l+1})^{+}\mathbf{X}_{k}^{v,l+1} + \mathbf{B}^{l+1}\mathbf{Y}_{k}^{v,l+1} + \mathbf{Z}_{k}^{v,l+1}  \bigg) \mathbf{H}_{k}^{H} .
\end{aligned}
\end{equation}
Thus, we have the recurrence relation for the gradient with respect to $\{ \mathbf{W}_{n} \}$ as
\begin{equation} \label{gradientGW}
\mathbf{G}_{n}^{w,l} = \sum\limits_{k=1}^K \textrm{Tr}( \mathbf{J}_{k}^{v,l}+\mathbf{L}_{k}^{v,l} )\frac{\sigma^{2}_{k}}{P_{T}}(\mathbf{U}_{n}^{l})^{H}\mathbf{U}_{n}^{l}
+ \sum\limits_{k=1}^K (\mathbf{U}_{n}^{l})^{H}\mathbf{H}_{n}( \mathbf{J}_{k}^{v,l}+\mathbf{L}_{k}^{v,l} )\mathbf{H}_{n}^{H}\mathbf{U}_{n}^{l} + \mathbf{G}_{n}^{v,l}\mathbf{M}_{n}^{v,l}\mathbf{U}_{n}^{l}.
\end{equation}
Similarly, we obtain the recurrence relation of the gradients with respect to $\{ \mathbf{U}_{n}, \mathbf{V}_{n} \}$ in adjacent layers as
\begin{subequations}  \label{recurrence}
\begin{eqnarray}
& &\tilde{\mathbf{G}}_{n}^{v,l} = - ( \mathbf{J}_{n}^{w,l+1} + \mathbf{Y}_{n}^{w,l+1}\mathbf{G}_{n}^{w,l+1} )(\mathbf{U}_{n}^{l+1})^{H}\mathbf{H}_{n} + \mathbf{M}^{u,l+1}_{n} \notag \\
& & \quad \quad \quad + \sum\limits_{k=1}^K\textrm{Tr}( \mathbf{N}_{k}^{u,l+1} )\frac{\sigma^{2}_{k}}{P_{T}}(\mathbf{V}_{n}^{l})^{H}
+ \sum\limits_{k=1}^K (\mathbf{V}_{n}^{l})^{H}\mathbf{H}_{k}^{H}( \mathbf{N}_{k}^{u,l+1} )\mathbf{H}_{k},  \\
& &\mathbf{G}_{n}^{u,l} = - \big( (\mathbf{J}_{n}^{w,l})^{H} + (\mathbf{G}_{n}^{w,l})^{H}(\mathbf{Y}_{n}^{w,l})^{H} \big)(\mathbf{V}_{n}^{l-1})^{H}\mathbf{H}_{n}^{H}
+ \sum\limits_{k=1}^K \mathbf{W}_{n}^{l}(\mathbf{U}_{n}^{l})^{H}\mathbf{H}_{n}( \mathbf{J}_{k}^{v,l}+\mathbf{L}_{k}^{v,l} )\mathbf{H}_{n}^{H}  \notag \\
& & \quad \quad \quad + \sum\limits_{k=1}^K \textrm{Tr}( \mathbf{J}_{k}^{v,l}+\mathbf{L}_{k}^{v,l} )\frac{\sigma^{2}_{k}}{P_{T}}\mathbf{W}_{n}^{l}(\mathbf{U}_{n}^{l})^{H}
+ \sum\limits_{k=1}^K (\mathbf{W}_{n}^{l})^{H}(\mathbf{U}_{n}^{l})^{H}\mathbf{H}_{n}( (\mathbf{J}_{k}^{v,l})^{H}+(\mathbf{L}_{k}^{v,l})^{H} )\mathbf{H}_{n}^{H} \notag \\
& & \quad \quad \quad + \sum\limits_{k=1}^K \textrm{Tr}( (\mathbf{J}_{k}^{v,l})^{H}+(\mathbf{L}_{k}^{v,l})^{H} )\frac{\sigma^{2}_{k}}{P_{T}}(\mathbf{W}_{n}^{l})^{H}(\mathbf{U}_{n}^{l})^{H}
+ \mathbf{W}_{n}^{l}\mathbf{G}_{n}^{v,l}\mathbf{M}^{v,l}_{n},
\end{eqnarray}
\end{subequations}
where
\begin{equation}  \label{networkJLM}
\begin{aligned}
& \mathbf{J}_{k}^{u,l+1} \triangleq \bigg( \mathbf{X}_{k}^{u,l+1}\mathbf{H}_{k}\mathbf{V}_{k}^{l}\mathbf{G}_{k}^{u,l+1} \bigg)\circ
\bigg( -(\mathbf{A}_{k}^{l})^{+}\circ (\mathbf{A}_{k}^{l})^{+} \bigg)^{T},    \\
& \mathbf{L}_{k}^{u,l+1} \triangleq \mathbf{Y}_{k}^{u,l+1}\mathbf{H}_{k}\mathbf{V}_{k}^{l}\mathbf{G}_{k}^{u,l+1} ,    \\
& \mathbf{M}_{k}^{u,l+1} \triangleq \mathbf{G}_{k}^{u,l+1} \bigg( (\mathbf{A}_{k}^{l})^{+}\mathbf{X}_{k}^{u,l+1} + \mathbf{A}_{k}^{l}\mathbf{Y}_{k}^{u,l+1}
+ \mathbf{Z}_{k}^{u,l+1}  \bigg) \mathbf{H}_{k} ,    \\
& \mathbf{N}_{k}^{u,l+1} \triangleq \mathbf{J}_{k}^{u,l+1}+\mathbf{L}_{k}^{u,l+1}+(\mathbf{J}_{k}^{u,l+1})^{H}+(\mathbf{L}_{k}^{u,l+1})^{H}, \\
& \mathbf{J}_{k}^{w,l+1} \triangleq \bigg( \mathbf{X}_{k}^{w,l+1}\mathbf{G}_{k}^{w,l+1} \bigg)\circ \bigg( -(\mathbf{E}_{k}^{l+1})^{+}\circ (\mathbf{E}_{k}^{l+1})^{+} \bigg)^{T}.
\end{aligned}
\end{equation}
We need to normalize each $\mathbf{V}_{k}$ by $P_{T}$ at the end of each layer, the gradient of $\mathbf{V}$ is given by
\begin{equation} \label{gradientGV}
\mathbf{G}_{n}^{v,l} = \tilde{\mathbf{G}}_{n}^{v,l} + \tilde{\mathbf{G}}_{n}^{v,l}\sqrt{P_{t}}a^{-\frac{1}{2}} - \sum\limits_{m=1}^K \frac{1}{2}\textrm{Tr}\bigg( \tilde{\mathbf{G}}_{m}^{v,l}\mathbf{V}_{m}^{l} + (\tilde{\mathbf{G}}_{m}^{v,l})^{H}(\mathbf{V}_{m}^{l})^{H} \bigg) \sqrt{P_{t}} a^{-\frac{3}{2}} (\mathbf{V}_{n}^{l})^{H},
\end{equation}
where $a\triangleq \sum\limits_{k=1}^K \textrm{Tr}(\mathbf{V}^{l}_{k}(\mathbf{V}_{k}^{l})^{H})$.

\section{ Details about \eqref{totaldiff}-\eqref{dW}  }
\label{Appendix C}
In this appendix, we show the details about \eqref{totaldiff}-\eqref{dW}. 
Recalling the expression of $\mathbf{W}_{k}^{l+1}$, $\mathbf{E}_{k}^{l+1}$, and $\mathbf{J}_{k}^{l+1}$ in \eqref{networkUWV}, \eqref{networkABE}, and \eqref{networkJLM}, respectively, and based on the results from Theorem \ref{theorem Chain rule} and Lemma \ref{Properties}, we have the following recurrence relation from $\mathbf{W}_{k}^{l+1}$ to $\mathbf{V}_{k}^{l}$ and $(\mathbf{U}_{k}^{l+1})^{H}$,
\begin{equation}
\begin{aligned}
\textrm{Tr}( \mathbf{G}_{k}^{w,l+1}d\mathbf{W}_{k}^{l+1} ) &= \textrm{Tr}\bigg\{  \mathbf{G}_{k}^{w,l+1} \bigg( -(\mathbf{E}_{k}^{l+1})^{+} \circ (\mathbf{E}_{k}^{l+1})^{+} \circ d\mathbf{E}_{k}^{l+1} \bigg) \mathbf{X}_{k}^{w,l+1} +  \mathbf{G}_{k}^{w,l+1}  d\mathbf{E}_{k}^{l+1}  \mathbf{Y}_{k}^{w,l+1} \bigg\}  \\
&= \textrm{Tr}\bigg\{ \big( \mathbf{J}_{k}^{w,l+1} + \mathbf{Y}_{k}^{w,l+1}\mathbf{G}_{k}^{w,l+1} \big) d\mathbf{E}_{k}^{l+1}  \bigg\}  \\
&= -\textrm{Tr}\bigg\{ \big( \mathbf{J}_{k}^{w,l+1} + \mathbf{Y}_{k}^{w,l+1}\mathbf{G}_{k}^{w,l+1} \big)(\mathbf{U}_{k}^{l+1})^{H}\mathbf{H}_{k} d\mathbf{V}_{k}^{l}   \\
&\quad \quad \quad \quad \quad \quad \quad
+  \mathbf{H}_{k}\mathbf{V}_{k}^{l}\big( \mathbf{J}_{k}^{w,l+1} + \mathbf{Y}_{k}^{w,l+1}\mathbf{G}_{k}^{w,l+1} \big)d(\mathbf{U}_{k}^{l+1})^{H} \bigg\}.
\end{aligned}
\end{equation}
The expression of $\textrm{Tr}( \mathbf{G}_{k}^{u,l+1}d\mathbf{U}_{k}^{l+1} )$ and $\textrm{Tr}( \mathbf{G}_{k}^{v,l+1}d\mathbf{V}_{k}^{l+1} )$ can be obtained similarly.
Then, we add these terms together and write them in the form of \eqref{totaldiff}. Finally, we retain the terms with $d\mathbf{W}_{k}^{l}$ and  obtain \eqref{dW} .

\end{appendices}

\bibliography{references}

\begin{thebibliography}{1}
\bibitem{MIMO01}
E. Björnson, L. Sanguinetti, J. Hoydis, and M. Debbah, ``Optimal design of energy-efficient multi-user MIMO systems: Is massive MIMO the answer?" \textit{IEEE Trans. Wireless Commun.}, vol. 14, no. 6, pp. 3059-3075, Jun. 2015.

\bibitem{MIMO02}
H. Q. Ngo, E. G. Larsson, and T. L. Marzetta, ``Energy and spectral efficiency of very large multiuser MIMO systems," \textit{IEEE Trans. Commun.}, vol. 61, no. 4, pp. 1436-1449, Apr. 2013.

\bibitem{MIMO1}
S. Ye and R. S. Blum, ``Optimized signaling for MIMO interference systems with feedback," \textit{IEEE Trans. Signal Process.}, vol. 51, no. 11, pp. 2839-2848, Nov. 2003.

\bibitem{MIMO2}
G. Scutari, D. P. Palomar, and S. Barbarossa, ``The MIMO iterative waterfilling algorithm," \textit{IEEE Trans. Signal Process.}, vol. 57, no. 5, pp. 1917-1935, May 2009.

\bibitem{MIMO3}
S. S. Christensen, R. Argawal, E. de Carvalho, and J. M. Cioffi,
``Weighted sum-rate maximization using weighted MMSE for MIMO-BC beamforming design," \textit{IEEE Trans. Wireless Commun.}, vol. 7, no. 12, pp. 1-7, Dec. 2008.

\bibitem{IWFA}
W. Yu and J. M. Cioffi, ``FDMA capacity of Gaussian multiple-access channel with ISI,"  \textit{IEEE Trans. Commun.}, vol. 50, no. 1, pp. 102-111, Jan. 2002.

\bibitem{SDR}
N. D. Sidiropoulos, T. N. Davidson, and Z. Luo, ``Transmit beamforming for physical-layer multicasting,"  \textit{IEEE Trans. Signal Process.}, vol. 54, no. 6, pp. 2239-2251, Jun. 2006.

\bibitem{WMMSE}
Q. Shi, M. Razaviyayn, Z. Luo, and C. He, ``An iteratively weighted MMSE approach to distributed sum-utility maximization for a MIMO interfering broadcast channel,"  \textit{IEEE Trans. Signal Process.}, vol. 59, no. 9, pp. 4331-4340, Sep. 2011.

\bibitem{Spectral}
Q. Shi and M. Hong, ``Spectral efficiency optimization for millimeter wave multiuser MIMO systems," \textit{IEEE J. Sel. Topics Signal Process.}, vol. 12, no. 3, pp. 455-468, Jun. 2018.


\bibitem{DL}
Y. LeCun, Y. Bengio, and G. Hinton, ``Deep learning,"  \textit{Nature}, vol. 521, no. 7553, pp. 436-444, 2015.

\bibitem{LearnOptimize}
H. Sun, X. Chen, Q. Shi, M. Hong, X. Fu, and N. D. Sidiropoulos, ``Learning to optimize: Training deep neural networks for interference management,"  \textit{IEEE Trans. Signal Process.}, vol. 66, no.20, pp. 5438-5453, Oct. 2018.

\bibitem{PowerControl}
W. Lee, M. Kim, and D.-H. Cho, ``Deep power control: Transmit power control scheme based on convolutional neural network," \textit{IEEE Commun. Lett.}, vol. 22, no. 6, pp. 1276-1279, Apr. 2018.

\bibitem{Towards}
F. Liang, C. Shen, W. Yu, and F. Wu, ``Towards optimal power control via ensembling deep neural networks," \textit{IEEE Trans. Commun.}, vol. 68, no. 3, pp. 1760-1776, Mar. 2020.

\bibitem{Spatial}
W. Cui, K. Shen, and W. Yu, ``Spatial deep learning for wireless scheduling," \textit{IEEE J. Sel. Areas Commun.}, vol. 37, no. 6, pp. 1248-1216, Jun. 2019.

\bibitem{Graph}
M. Lee, G. Yu, and G. Y. Li, ``Graph embedding based wireless link scheduling with few training samples," \emph{arXiv preprint arXiv:1906.02871}, 2019.

\bibitem{Channelesti}
H. He, C. Wen, S. Jin, and G. Y. Li, ``Deep learning-based channel estimation for beamspace mmWave massive MIMO systems," \textit{IEEE Wireless Commun. Lett.}, vol. 7, no. 5, pp. 852-855, Oct. 2018.

\bibitem{CSIfeedback}
Z. Liu, L. Zhang, and Z. Ding, ``Exploiting bi-directional channel reciprocity in deep learning for low rate massive MIMO CSI feedback," \textit{IEEE Wireless Commun. Lett.}, vol. 8, no. 3, pp. 889-892, Jun. 2019.


\bibitem{DeepUnfold}
J. R. Hershey, J. L. Roux, and F. Weninger, ``Deep unfolding: Model-based inspiration of novel deep architectures,"
\emph{arXiv preprint arXiv:1904.03406}, 2014.


\bibitem{TopicModel}
J. Chien and C. Lee, ``Deep unfolding for topic models,"  \textit{IEEE Trans. Pattern Anal. Mach. Intell.}, vol. 40, no. 2, pp. 318-331, Feb. 2018.

\bibitem{DeepProximal}
R. Liu, S. Cheng, L. Ma, X. Fan, and Z. Luo, ``Deep proximal unrolling: Algorithmic framework, convergence analysis and applications,"  \textit{IEEE Trans. Image Process.}, vol. 28, no. 10,
pp. 5013-5026, Oct. 2019.

\bibitem{Interpret}
V. Monga, Y. Li, and Y. C. Eldar, ``Algorithm unrolling: Interpretable, efficient deep learning for signal and image processing,"  \emph{arXiv preprint arXiv:1912.10557}, 2019.


\bibitem{RealTime}
L. Zhang, G. Wang, and G. B. Giannakis, ``Real-time power system state estimation and forecasting via deep unrolled neural networks,"  \textit{IEEE Trans. Signal Process.}, vol. 67, no. 15, pp. 4069-4077, Aug. 2019.

\bibitem{UnfoldSurvey}
A. B. Stimming and C. Studer, ``Deep unfolding for communications systems: A survey and some new directions," in  \textit{IEEE Int. Workshop on Signal Process. Systems (SiPS)}, Oct. 2019, pp. 266-271.

\bibitem{Detect1}
N. Samuel, T. Diskin, and A. Wiesel, ``Learning to detect,"  \textit{IEEE Trans. Signal Process.}, vol. 67, no. 10, pp. 2554-2564, May 2019.

\bibitem{Detect2}
H. He, C.-K. Wen, S. Jin, and G. Y. Li, ``A model-driven deep learning
network for MIMO detection," in \textit{IEEE Global Conference on Signal and Information Processing (GlobalSIP)}, Nov. 2018, pp. 584-588.

\bibitem{ISTA}
K. Gregor and Y. LeCun, ``Learning fast approximations of sparse coding," in \textit{Proc. 27th Int. Conf. Mach. Learn.}, 2010, pp. 399-406.

\bibitem{Coding}
S. Cammerer, T. Gruber, J. Hoydis, and S. ten Brink, ``Scaling deep
learning-based decoding of polar codes via partitioning," in \textit{IEEE Global
Communications Conference (GLOBECOM)}, Dec. 2017, pp. 1-6.

\bibitem{AMP}
M. Borgerding, P. Schniter, and S. Rangan, ``AMP-Inspired deep networks for sparse linear inverse problems,"  \textit{IEEE Trans. Signal Process.}, vol. 65, no. 16, pp. 4293-4308, Aug. 2017.

\bibitem{LearnResource}
M. Eisen, C. Zhang, L. F. O. Chamon, and D. D. Lee, ``Learning optimal resource allocations in wireless systems," \textit{IEEE Trans. Signal Process.}, vol. 67, no. 10, pp. 2775-2790, May 2019.


\bibitem{Appli}
H. Lee, S. Lee, and T. Q. S. Quek, ``Deep learning for distributed optimization: Applications to wireless resource management,"  \textit{IEEE J. Sel. Areas Commun.}, vol. 37, no. 10, pp. 2251-2266, Oct. 2019.


\bibitem{ALiu}
A. Liu, V. Lau, and M. Zhao, ``Online successive convex approximation for two-stage stochastic nonconvex optimization,"
\textit{IEEE Trans. Signal Process.}, vol. 66, no. 22, pp. 5941-5955, Nov. 2018.



\bibitem{MatrixDiff}
K. Brandt and M. S. Pedersen, \textit{The Matrix Cookbook}. Technical University of Denmark 7.15 (2008): 510.

\end{thebibliography}

\end{document}